\documentclass[prd,aps,preprint,tightenlines,superscriptaddress,nofootinbib,showpacs,showkeys]{revtex4}
\usepackage{mathrsfs}
\usepackage{amsfonts}
\usepackage{array}
\usepackage{epsfig}

\usepackage{amsmath}    
\usepackage{verbatim}   
\usepackage{color}      
\usepackage{colordvi}

\usepackage{subfigure}  
\usepackage{hyperref}   

\usepackage{pstricks,rotating, graphicx}

\graphicspath{{../pics/}}
\DeclareGraphicsExtensions{.eps,.ps}

\usepackage{pslatex}
\usepackage{txfonts}
\usepackage[latin1]{inputenc}
\usepackage[T1]{fontenc}

\def\muf{{\mu^{}_f}}
\def\mufs{{\mu^{\,2}_f}}
\def\mur{{\mu^{}_r}}
\def\murs{{\mu^{\,2}_r}}

\def\mbar{\overline{m}}
\def\mmu{m(\mur)}

\newcommand{\msbar}{$\overline{\mathrm{MS}}\, $}

\begin{document}

\begin{titlepage}
\thispagestyle{empty}
\noindent
DESY 14-055\\
DO-TH 14/06\\
LPN 14-063\\
SFB/CPP-14-21\\
MITP/14-031
\hfill
April 2014 \\
\vspace{1.0cm}

\begin{center}
  {\bf
    \Large Determination of Strange Sea Quark Distributions \\ 
           from Fixed-target and Collider Data\\
  }
  \vspace{1.25cm}
 {\large
   S.~Alekhin$^{\, a,b}$,
   J.~Bl\"umlein$^{\, a}$,
   L.~Caminada$^{\, c}$,
   K.~Lipka$^{\, d}$,
   K.~Lohwasser$^{\, a}$,
   \\[0.5ex]
   S.~Moch$^{\, a,e}$,
   R.~Petti$^{\, f}$,
   and 
   R.~Pla\v cakyt\. e$^{\, d}$
   \\
 }
 \vspace{1.25cm}
 {\it
   $^a$Deutsches Elektronensynchrotron DESY \\
   Platanenallee 6, D--15738 Zeuthen, Germany \\
   \vspace{0.2cm}
   $^b$Institute for High Energy Physics \\
   142281 Protvino, Moscow region, Russia\\
   \vspace{0.2cm}
   $^c$Physik Institut, Universit\"{a}t Z\"{u}rich \\
   Winterthurerstra{\ss}e 190, CH--8057 Z\"{u}rich, Switzerland \\
   \vspace{0.2cm}
   $^d$Deutsches Elektronensynchrotron DESY \\
   Notkestra{\ss}e 85, D--22607 Hamburg, Germany \\
   \vspace{0.2cm}
   $^e$ II. Institut f\"ur Theoretische Physik, Universit\"at Hamburg \\
   Luruper Chaussee 149, D--22761 Hamburg, Germany \\
   \vspace{0.2cm}
   $^f$ Department of Physics and Astronomy, University of South Carolina \\
   712 Main Street, Columbia, SC 29208, USA \\
 }
  \vspace{1.4cm}
  \large {\bf Abstract}
  \vspace{-0.2cm}
\end{center}
We present an improved determination of the strange sea distribution 
in the nucleon with constraints coming from the recent charm production data in 
neutrino-nucleon deep-inelastic scattering by the NOMAD and CHORUS experiments 
and from charged current inclusive deep-inelastic scattering at HERA.
We demonstrate that the results are consistent with the data from the ATLAS and the CMS 
experiments on the associated production of $W^\pm$-bosons with $c$-quarks. 
We also discuss issues related to the recent strange sea determination by the 
ATLAS experiment using LHC collider data.  
\end{titlepage}

\newpage
\setcounter{footnote}{0}
\setcounter{page}{1}

\section{Introduction}

The accurate knowledge of the flavor decomposition of the quark distributions
in the proton is an important prerequisite in Quantum Chromodynamics (QCD) 
for precision phenomenology at current colliders.
While the individual valence and sea parts of the $u$- and $d$-quark parton distribution functions (PDFs)
are relatively well constrained by existing data, the strange sea in the proton is only poorly known.
Nevertheless, the $s$-quark PDF affects predictions for the cross-sections of a
significant number of hadron processes. 
These include in particular the processes involving 
the exchange of a $W$-boson with space- or time-like kinematics, such
as charm production in the charged-current (CC) deep-inelastic
scattering (DIS) and $W$-boson production in association with charm quarks
at proton-proton colliders.
In particular, the knowledge of the strange and anti-strange sea quark PDFs, as well as of the related charm 
quark production in CC DIS interactions, are the dominant source of uncertainty in 
probing precision electroweak physics with (anti)neutrinos~\cite{Zeller:2001hh}.  

For a long time, the information on the strange sea quark content of the nucleon has almost 
entirely relied on the data from charm di-muon production in (anti)neutrino-iron DIS interactions 
by the NuTeV and CCFR experiments~\cite{Goncharov:2001qe}.   
These data have a kinematic coverage in the parton momentum fraction $x$ and the virtuality $Q^2$ 
limited by the fixed target kinematics and by the (anti)neutrino beam energy.  
It must be noted, however, that charm di-muon data from $\nu(\bar \nu)$-nucleon DIS interactions 
also depend on the knowledge of the semi-leptonic branching ratio $B_\mu$ for the inclusive decays of 
different charmed hadrons into muons. 

The situation has significantly improved with the recent publication of new data samples from 
different experiments. The new precision measurement of charm di-muon production in neutrino-iron 
DIS interactions by NOMAD~\cite{Samoylov:2013xoa} has substantially reduced both experimental 
and model uncertainties with respect to the NuTeV and CCFR measurements. In addition, the CHORUS 
experiment~\cite{KayisTopaksu:2011mx} has released a new direct measurement of inclusive charm 
production in nuclear emulsions insensitive to $B_\mu$.    
Furthermore, improved theoretical descriptions in perturbative QCD for some of
the underlying hard scattering reactions have become available. 
Specifically, the perturbative QCD corrections at next-to-next-to-leading order (NNLO) in QCD 
for charm quark production in CC DIS can be applied, which describe the heavy-quark coefficient functions 
at asymptotic values of 
$Q^2 \gg m_{h}^2$~\cite{Buza:1997mg,moch:2013cc,Blumlein:2014fqa}, where
$m_{h}$ is the mass of heavy quark, $c$ or $b$. 
With the Wilson coefficients in this asymptotic regime it becomes possible 
to include consistently HERA cross-section measurements for CC inclusive DIS~\cite{Aaron:2009aa} at NNLO QCD into 
the analysis. Those data provide additional constraints on the $s$-quark PDF in a much wider kinematical 
range for $Q^2$.

At the Large Hadron Collider (LHC) measurements of $W^\pm$-boson production, either inclusive or
associated with a $c$-quark jet or $D^*$-meson allow for tests of those strange sea determinations from DIS data.
The LHC measurements of inclusive $W^\pm$ and $Z$-boson production can be used in a QCD analysis at NNLO with the 
theory predictions available fully differentially in the gauge-boson kinematics. ATLAS has reported an enhancement of  
the $s$-quark distribution~\cite{Aad:2012sb} with respect to other measurements. For the exclusive 
process $pp \to W^\pm + c$ the QCD corrections are only known to next-to-leading order (NLO), which 
implies larger theoretical uncertainties. 
Nevertheless, the available $W+c$ data by CMS~\cite{Chatrchyan:2013uja} and ATLAS~\cite{Aad:2014xca}
offer valuable insight and allow for cross checks, 
both of the above mentioned $epWZ$-fit by ATLAS~\cite{Aad:2012sb} based on electron-proton DIS and inclusive $W/Z$-boson data
as well as of strange sea determinations from global PDF fits.

This paper is organized as follows. 
Sec.~\ref{sec:analysis} introduces the framework of the analysis, 
which is based on a global fit of PDFs 
by ABM~\cite{Alekhin:2009ni,Alekhin:2012ig,Alekhin:2013nda} 
using the world DIS data and the measurements of gauge-boson production from fixed targets and the LHC.
Sec.~\ref{sec:analysis} also summarizes briefly the new improvements in QCD theory for CC DIS charm quark production 
with a running $c$-quark mass.
Sec.~\ref{sec:datasets} gives a brief description of the new data sets 
from CHORUS, NOMAD~\cite{Goncharov:2001qe,KayisTopaksu:2011mx,Samoylov:2013xoa} 
and the LHC $W+c$ data~\cite{Chatrchyan:2013uja,Aad:2014xca}.
Sec.~\ref{sec:results} contains the results of the analysis 
including a new study of the energy dependence of the semi-leptonic branching ratio $B_\mu$ of charmed hadrons.
Starting from a fit to the combined data of NuTeV, CCFR, CHORUS and NOMAD~\cite{Goncharov:2001qe,KayisTopaksu:2011mx,Samoylov:2013xoa} 
the impact of individual data sets is quantified and the resulting shifts in the strange quark distributions are documented.
Particular care is also taken to control potential correlations with the $u$- and $d$-quark sea distributions.
In Sec.~\ref{sec:discussion} we compare the results with earlier determinations of the $s$-quark PDF. 
In particular, we comment on issues in the ATLAS determination of the
strange sea in the $epWZ$-fit~\cite{Aad:2012sb}
as well as in the $s$-quark PDF of NNPDF (version 2.3) obtained with a fit to the LHC data~\cite{Ball:2012cx}.
We conclude in Sec.~\ref{sec:conclusion}.

\renewcommand{\theequation}{\thesection.\arabic{equation}}
\setcounter{equation}{0}
\renewcommand{\thefigure}{\thesection.\arabic{figure}}
\setcounter{figure}{0}
\renewcommand{\thetable}{\thesection.\arabic{table}}
\setcounter{table}{0}
\section{Analysis framework}
\label{sec:analysis}

\subsection{Description of the global fit}

The present study is an extension of the ABM PDF fit which is based on the combination 
of the world DIS data and data for the Drell-Yan process obtained at
fixed-target and collider experiments (cf.~\cite{Alekhin:2013nda} and references therein). 
The flavor separation of  the $u$- and $d$-quark distributions in the nucleon is obtained 
with a good accuracy from the combination of proton and deuteron fixed-target data 
with the LHC data on $W$- and $Z$-boson production.
However, this approach can only provide a rather poor determination of the $s$-quark distribution, 
mainly due to the correlations with the $d$-quark distribution. 
Therefore, in the ABM fit the strange sea distribution is basically constrained 
by the data on charm di-muon production from neutrino-nucleus DIS, 
which constitutes a direct probe of the strangeness content of the nucleon~\cite{Alekhin:2008mb}.
An additional, though minor, constraint comes from the CC data obtained at HERA~\cite{Aaron:2009aa}. 
In line with the ABM12 fit, the $s$-quark sea distribution can be parameterized 
in a rather simple form at the initial scale for the PDF evolution $\mu_0=3~{\rm GeV}$: 
\begin{equation}
s(x,\mu_0)=A_s x^{a_{s}}(1-x)^{b_{s}}\, .
\end{equation}
where $A_s, a_s$ and $b_s$ are fitted parameters. 
This simple parameterization is sufficient to achieve a good description of the data. 

To a good approximation the present analysis is performed at the NNLO accuracy in QCD, 
i.e., by taking into account the NNLO corrections to the PDF evolution and to the
Wilson coefficients of the hard scattering processes. 
The description of the DIS data employs the three-flavor factorization scheme 
with the heavy $c$- and $b$-quarks appearing in the final state only. 
This approach provides a good agreement with the existing inclusive neutral-current (NC) DIS 
data up to the highest momentum transfer $Q^2$ covered by HERA~\cite{Alekhin:2013nda}. 
In particular, such an agreement is related to the use 
of the massive NNLO corrections for the NC heavy-quark production, 
along with the \msbar definition of the heavy-quark masses~\cite{Alekhin:2010sv,Alekhin:2012un}.
Instead, the CC DIS heavy-quark production has been described in the ABM12 fit with account of
the NLO corrections~\cite{Gottschalk:1980rv,Gluck:1996ve,Blumlein:2011zu} only. 
As a matter of fact, this approximation is adequate for the description of 
existing CC HERA data, in view of their relatively poor accuracy. 
However, for consistency, in the present analysis we also consider 
those NNLO QCD corrections which are applicable 
to CC DIS in the asymptotic region of $Q^2\gg m_{c}^2$. 

\subsection{Improved treatment of the massive charged-current NNLO corrections}
\label{sec:nnlo}

For CC DIS heavy-quark (charm) production at parton level proceeds in Born approximation 
in a $2 \to 1$ reaction as
\begin{equation}
\label{eq:ccborn}
s(p) + W^*(q) \to  c\, ,
\end{equation}
where $p$ and $q$ denote the momenta of the incoming $s$-quark and the off-shell
$W$-boson and define the well-known kinematical variables, Bjorken $x$ and $Q^2$, 
as $Q^2 = -q^2 >  0$ and $x = Q^2 /(2 p \cdot q)$~\footnote{At higher orders
  also $c\bar{c}$ pair production contributes~\cite{Buza:1997mg,Blumlein:2014fqa}.}.
The mass of the $s$-quark is neglected, the final state $c$-quark is heavy
and the coupling to the $W$-boson involves the usual parameters 
of the Cabibbo-Kobayashi-Maskawa (CKM) matrix.

The cross section is usually parameterized in terms of 
the heavy-quark DIS structure functions $F_k$, $k=1,2,3$, which depend on 
$x$, $Q^2$ and the heavy-quark mass $m_{c}$, 
and which can be written in the standard factorization approach to perturbative QCD as
\begin{equation}
  \label{eq:totalF2c}
  F_k(x,Q^2,m^2) =
  \sum\limits_{i = q,{\bar{q}},g} \,\,
  \int\limits_{\chi}^1\,
  {dz \over z} \,\, f_{i}\left({x \over z}, \mufs \right)\,\,
  {\cal C}_{k, i}\left(z,\xi,\murs,\mufs \right)
  \, ,
\end{equation}
i.e., as a convolution of PDFs $f_{i}$ and coefficient functions ${\cal C}_{k, i}$.
The scales for renormalization and factorization are $\mur$ and $\muf$ 
and the integration range over the parton momentum fraction $z$ is bounded by $\chi = x/\lambda$.
The kinematical variables $\xi$ in Eq.~(\ref{eq:totalF2c}) and $\lambda$ 
are given as $\xi = Q^2/m_{c}^2$ and 
$\lambda = 1/(1 +   m_{c}^2/Q^2) =  \xi/(1 + \xi)$, respectively.

The coefficient functions ${\cal C}_{k, i}$ of the hard parton scattering process in Eq.~(\ref{eq:totalF2c}) 
can be computed in a perturbative expansion in the strong coupling constant $\alpha_s = \alpha_s(\mur)$, 
\begin{equation}
  \label{eq:partonCcc-exp}
  {\cal C}_{k, i}(z,\xi,\murs,\mufs) =
  {\cal C}_{k, i}^{(0)} +  \alpha_s\, {\cal C}_{k, i}^{(1)} +  \alpha_s^2\, {\cal C}_{k, i}^{(2)}
  \, ,
\end{equation}
where ${\cal C}_{k, q}^{(0)} \simeq \delta(1-z)$ (up to the CKM parameters) and ${\cal C}_{k, g}^{(0)} = 0$ for $k=1,2,3$ due to Eq.~(\ref{eq:ccborn}).
For ${\cal C}_{k, i}^{(1)}$ the exact expressions are given in Refs.~\cite{Gottschalk:1980rv,Gluck:1996ve,Blumlein:2011zu} 
whereas for ${\cal C}_{k, i}^{(2)}$ results at asymptotic values of $Q^2 \gg m_{c}^2$ have been derived in
Refs.~\cite{Buza:1997mg,moch:2013cc,Blumlein:2014fqa}.
The former results have been derived for heavy-quark masses in the on-shell
renormalization~\footnote{The analytic relations for change in the 
asymptotic Wilson
  coefficients and massive operator
  matrix elements from the on-shell mass scheme to the \msbar scheme to
  $O(\alpha_s^3)$ were given in Ref.~\cite{Bierenbaum:2009mv}.}.
%
%
%
For consistency with the set-up of the ABM fit~\cite{Alekhin:2012ig,Alekhin:2013nda} 
which uses the \msbar definition for heavy-quark masses in DIS~\cite{Alekhin:2010sv,Alekhin:2012un} 
for an improved convergence and better stability under scale variation, we briefly summarize
below those changes from the pole mass scheme to the \msbar scheme which are necessary 
if the NNLO Wilson coefficients ${\cal C}_{k, i}^{(2)}$ at asymptotically large values  $\xi=Q^2/m_{c}^2$ are included.

The conversion uses the well-known relation between the pole mass $m_{c}$ and the running mass $\mmu$ in the \msbar scheme
\begin{equation}
  \label{eq:mpole-mbar}
  m_{c} = \mmu \* \left(1 + \alpha_s(\mur) d^{(1)}(\mur) + \alpha_s(\mur)^2 d^{(2)}(\mur) + \dots \right)
  \, ,
\end{equation}
where the coefficients $d^{(l)}$ are actually known to three-loop order~\cite{Gray:1990yh,Chetyrkin:1999qi,Melnikov:2000qh}.

We will derive explicit formulae through NNLO for the dependence of the structure functions on the \msbar mass $m_c(m_c)$.
In doing so, we extend the approach of Ref.~\cite{Alekhin:2010sv} to NNLO for CC DIS 
(see also~\cite{Langenfeld:2009wd,Aliev:2010zk} for the hadro-production of top-quarks pairs).
For the pole mass $m$ we have (suppressing all other arguments),
\begin{equation}
  \label{eq:F2c-mpole}
  F_k(m) =
  \alpha_s\, F_k^{(0)}(m) +  \alpha_s^2\, F_k^{(1)}(m) +
  \alpha_s^3\, F_k^{(2)}(m) 
  \, ,
\end{equation}
which is converted with the help of Eq.~(\ref{eq:mpole-mbar}) to the \msbar mass
$m_c(m_c)$ (for simplicity abbreviated as $\mbar$) according to 
\begin{eqnarray}
  \label{eq:F2c-mbar}
  F_k(\mbar) &=&
  \alpha_s\, F_k^{(0)}(\mbar) 
\\
& &
\nonumber
  + 
  \alpha_s^2\, \left( 
    F_k^{(1)}(\mbar) 
    + \mbar\, d^{(1)} \partial_m F_k^{(0)}(m) \biggr|_{m=\mbar}
  \right) 
\\
& &
\nonumber
  +
  \alpha_s^3\, \left( 
     F_k^{(2)}(\mbar) 
     + \mbar\, d^{(2)} \partial_m F_k^{(0)}(m) \biggr|_{m=\mbar}
     + \mbar\, d^{(1)} \partial_m F_k^{(1)}(m) \biggr|_{m=\mbar}
\right.
\\
& &
\nonumber
\left.
\hspace*{60mm}
     + {1 \over 2}\, \left(\mbar\, d^{(1)}\right)^2 \partial_m^2 F_k^{(0)}(m) \biggr|_{m=\mbar}
   \right) 
  \, ,
\end{eqnarray}
where the coefficients $d^{(l)}$ have to be evaluated for $\mur = \mbar$ (corresponding to the scale of $\alpha_s$).
Up to NLO, the necessary term, $\partial_m F_k^{(0)}(m)$,
is given in Ref.~\cite{Alekhin:2010sv} and the additional contributions at NNLO
can be obtained along the same lines.
As the current analysis is restricted to asymptotically large values $\xi=Q^2/m_{c}^2$ at NNLO,
the changes of the NNLO Wilson coefficients ${\cal C}_{k, i}^{(2)}$ 
from the pole mass scheme to the \msbar scheme need to be accounted for only to logarithmic accuracy in $\xi$.
To that end, it suffices to note that at order $\alpha_s^3$ in Eq.~(\ref{eq:F2c-mbar})
the second and the fourth term vanish for large $\xi$ as 
\begin{eqnarray}
  \label{eq:F2cc-mbar-dF0-asy}
  \mbar\, d^{(2)} \partial_m F_k^{(0)}(m) \biggr|_{m=\mbar} 
&\sim&
  \left(\mbar\, d^{(1)}\right)^2 \partial_m^2 F_k^{(0)}(m) \biggr|_{m=\mbar} 
\,\,\sim\,\,\,
{\cal O}\left(\xi^{-1}\right)
  \, ,
\end{eqnarray}
and, therefore can be neglected in the current approximation. 

In the third term at order $\alpha_s^3$ in Eq.~(\ref{eq:F2c-mbar}), i.e., in $\mbar\, d^{(1)} \partial_m F_k^{(1)}(m)$, 
only the Wilson coefficient for the gluon channel ${\cal C}_{k, g}^{(1)}$ contributes, 
since asymptotically the collinear singularity 
in ${\cal C}_{k, g}^{(1)} \sim \pm P_{qg}^{(0)} \ln(Q^2/m^2)$ is proportional
to the one-loop splitting function $P_{qg}^{(0)}$.
Therefore, for large $\xi$ the following replacement in the asymptotics at order $\alpha_s^2$ suffices,
\begin{eqnarray}
  \label{eq:lnreplace}
\alpha_s^2\, \ln^k\left(\xi\right) 
& \to &
\alpha_s^2\, \ln^k\left({Q^2 \over \mbar^2(1 + \alpha_s d^{(1)})}\right)
\simeq
\alpha_s^2\, \ln^k\left({Q^2 \over \mbar^2}\right) 
- \alpha_s^3\, k\, d^{(1)} \ln^{k-1}\left({Q^2 \over \mbar^2}\right) 
+ \dots
  \, ,
\end{eqnarray}
in order to generate the order $\alpha_s^3$ contribution in Eq.~(\ref{eq:F2c-mbar}) to the required accuracy.
All other contributions, especially the boundary terms from $\partial_m \chi$ in Eq.~(\ref{eq:totalF2c}) 
vanish as ${\cal O}\left(\xi^{-1}\right)$. 

Experience from the case of NC DIS shows that the asymptotic expansion in powers of $\ln^k(\xi)$ 
agrees well with the exact result for the full mass dependence already at moderate values of $\xi \gtrsim 10$.
For the CC DIS case, the validity can be checked at NLO 
with the respective expressions for the known Wilson coefficients, i.e., by comparing exact versus asymptotic.
Such comparison reveals that starting from values of $\xi \gtrsim 50$ 
(depending on the $x$-values not being too large) the asymptotic expressions for the individual
channels (${\cal C}_{k, g}^{(1)}$ and ${\cal C}_{k, q}^{(1)}$) reproduce the exact results 
to better than ${\cal O}(10 \dots 20 \%)$, but mostly to much better accuracy. 
Asymptotically, for $\xi$ very large, agreement within a few percent is reached.
For larger $x$-values, $x \ge 0.1$ the on-set of the asymptotic behavior is
generally delayed due to the numerical dominance of threshold Sudakov logarithms, which
can be resummed to all orders in perturbation theory, see~\cite{Corcella:2003ib}.
By combing those result on threshold logarithms with the asymptotic expressions one could, in principle, 
arrive at further refinements of the NNLO approximation for the CC DIS Wilson
coefficients along the lines of Ref.~\cite{Kawamura:2012cr} for NC DIS.
Given the overall small numerical size of the higher order CC DIS QCD corrections, 
as well as the accuracy and kinematical coverage of the existing experimental
data, we leave this task for future studies.

\begin{figure}[tbh]
\centerline{
\includegraphics[width=14cm]{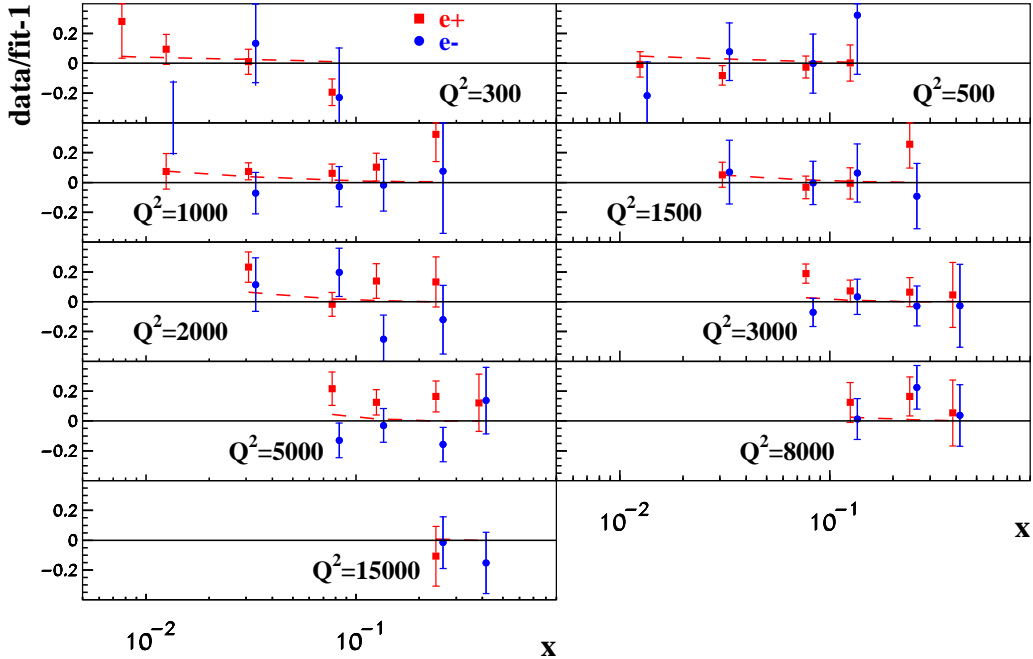}}
  \caption{\small
    \label{fig:heracc}
Pulls with respect to the ABM12
PDF fit~\cite{Alekhin:2013nda} for the HERA CC inclusive DIS 
cross section data of Ref.~\cite{Aaron:2009aa} in different bins 
of the momentum transfer $Q^2$ (squares: positron beam; 
circles: electron beam). The dashes display the impact of the NNLO
corrections to the CC massive Wilson 
coefficients~\cite{Buza:1997mg,moch:2013cc,Blumlein:2014fqa}  
derived with the \msbar $c$-quark mass
definition on the $e^+$-initiated CC cross sections.
}
\end{figure}

Fig.~\ref{fig:heracc} displays the comparison of the inclusive CC DIS cross
section data from HERA~\cite{Aaron:2009aa} with the NNLO QCD corrections as
discussed here in the text and using the \msbar $c$-quark mass definition with
$m_c(m_c) = 1.24$~GeV (see the ABM12 PDF fit~\cite{Alekhin:2013nda}).
The absolute magnitude of the NNLO corrections (taken at the nominal scale $\mu^2 = Q^2 + m_c(m_c)^2$)
is small, so that the main virtue of NNLO accuracy 
lies in an overall reduction of the scale uncertainty~\footnote{Main
  heavy-quark CC DIS corrections to 
NNLO~\cite{Buza:1997mg,moch:2013cc,Blumlein:2014fqa} used in our analysis 
are included into version 2.0 of the {\tt OPENQCDRAD} code and are publicly available 
online~\cite{openqcdrad:2014}. The code of Ref.~\cite{Blumlein:2014fqa}
including all Wilson coefficients is available on request to Johannes.Bluemlein@desy.de.
}. 

\renewcommand{\theequation}{\thesection.\arabic{equation}}
\setcounter{equation}{0}
\renewcommand{\thefigure}{\thesection.\arabic{figure}}
\setcounter{figure}{0}
\renewcommand{\thetable}{\thesection.\arabic{table}}
\setcounter{table}{0}
\section{New data sets relevant for the strange sea determination} 
\label{sec:datasets}

\subsection{Charm di-muon production in $\nu$-iron DIS}
\label{sec:NOMAD} 

Charm production in $\nu(\bar \nu)$N DIS interactions offers ideally the most direct probe 
of strange sea quark distributions. The most common experimental technique is to detect 
inclusive semi-leptonic decays of charmed hadrons into muons, resulting in a clean signature 
with two muons of opposite charge in the final state.    
The di-muon production cross section depends on the inclusive semi-leptonic branching 
ratio $B_\mu$, which reads
\begin{equation}
B_\mu=\sum_h B^h_\mu f_h(E_\nu)\, , 
\end{equation}
where $B^h_\mu$ are the semi-leptonic branching ratios of the individual charmed hadrons, $h=D^0,D^+,D_s,\Lambda_c$ 
(where the $\Lambda_c$ notation includes heavier charmed baryons), produced in the 
neutrino-nucleon collisions, $f_h(E_\nu)$ are the corresponding production fractions,  
and $E_\nu$ is the neutrino beam energy. 
In general, the charmed fractions $f_h$ depend on the incoming neutrino energy.
This fact can be explained by the contributions from quasi-elastic $\Lambda_c$
and diffractive $D_s^\pm$ production.
These two contributions are significant mainly at low energies and they do not affect the value 
of $B_\mu$ at $E_\nu \gtrsim 50~{\rm GeV}$, where the spectrum flattens out. 

\begin{figure}[tbh]
\centerline{
  \includegraphics[width=10cm]{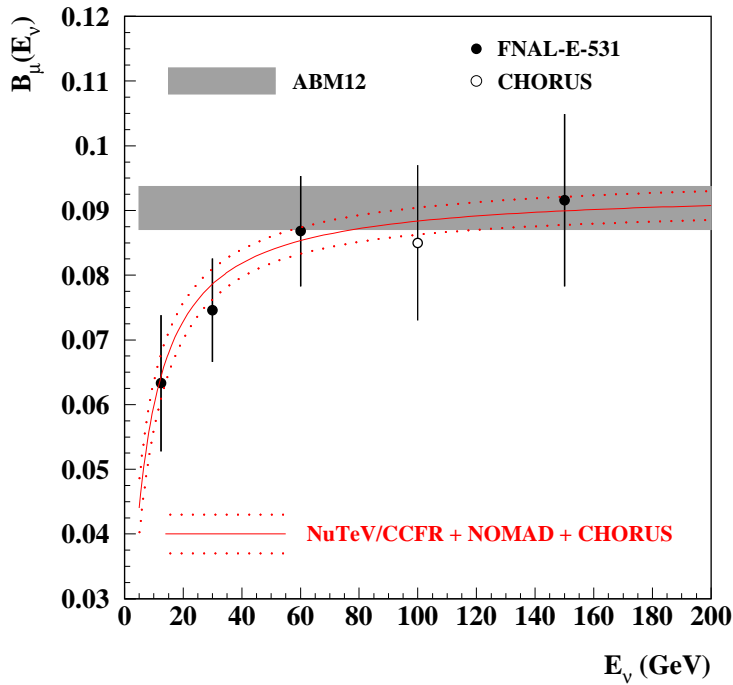}}
  \caption{\small
    \label{fig:bmu}
    Value of the semi-leptonic branching ratio 
    $B_\mu(E_\nu)$ obtained in the variant of the present analysis with the 
    combination of the NOMAD~\cite{Samoylov:2013xoa} and 
    CHORUS~\cite{KayisTopaksu:2011mx} data added (solid line: central value, dots: $1\sigma$ error band), 
    compared with the corresponding $1\sigma$ band obtained in the
    ABM12 fit~\cite{Alekhin:2013nda} (shaded area). 
    The measurements of $B_\mu$ by the emulsion experiments 
    FNAL-E-531~\cite{Ushida:1988rt}
    (full circles) and CHORUS~\cite{KayisTopaksu:2005je} 
    (hollow circles) are given for comparison. 
}
\end{figure}

Typically, a minimal energy threshold is required for the muons to be identified
experimentally, in order to suppress the background from $\pi,K$ muonic decays.
This experimental requirement results in an acceptance correction for the 
undetected phase space, which enhance the sensitivity of the charm di-muon measurements to 
the charm quark fragmentation into hadrons. A second potential source of uncertainty is 
related to the use of heavy nuclei as (anti)neutrino target, resulting in nuclear 
modifications on the measured cross-sections.  
In this paper we consider three charm di-muon data sets obtained on Fe-targets by the 
NuTeV, CCFR~\cite{Goncharov:2001qe} and NOMAD~\cite{Samoylov:2013xoa} experiments. 

The NuTeV and CCFR data samples~\cite{Goncharov:2001qe} --- corresponding to 5102 (1458) and 5030 (1060) 
$\nu(\bar \nu)$-induced di-muon events, respectively --- have been the only charm di-muon data used in earlier 
fits~\cite{Alekhin:2008mb,Alekhin:2012ig}, providing most of the information on the strange 
sea quark distributions. Both experiments have measured the absolute differential cross-section 
for charm di-muon production on iron, $d \sigma^2_{\mu \mu} / dxdy$.  
The minimal energy threshold for the muon detection was 5~GeV. 
It is worth noting that neglecting the dependence of $B_\mu$ on $E_\nu$ has been a good 
approximation in the analysis of the high energy NuTeV/CCFR di-muon data~\cite{Alekhin:2008mb}.

The new measurement of charm di-muon production in $\nu$-Fe interactions 
by the NOMAD experiment~\cite{Samoylov:2013xoa} is characterized by an increase by a factor of three in the statistics 
(15344 events), as well as by an improved understanding of the uncertainties discussed above. 
The minimal energy threshold for the muon detection was lowered to 3~GeV, allowing for a 
substantial increase of the detected phase space, thus reducing the sensitivity to the charm quark 
fragmentation. The NOMAD data extends down to $E_\nu=6$~GeV, providing a better sensitivity 
to charm production parameters close to the $m_c$ threshold.  
The NOMAD experiment measured the ratio ${\cal R}_{\mu \mu} \equiv \sigma_{\mu\mu} / \sigma_{\rm CC}$ 
between the charm di-muon cross-section and the inclusive charged current
cross-section (two muons versus a single muon) 
as a function of the three independent variables $E_\nu$, $x$, and 
the partonic center-of-mass energy $\sqrt{\hat{s}}$, for $Q^2>1$ GeV$^2$.  
This ratio offers a large cancellation of both experimental and theoretical uncertainties, 
including the nuclear corrections related to the Fe-target~\cite{Kulagin:2004ie,Kulagin:2007ju,Kulagin:2010gd}. 

In the energy region covered by the NOMAD data the inclusive semi-leptonic branching ratio 
$B_{\mu}$ demonstrates a clear dependence on $E_{\nu}$. To account for this dependence we parameterize 
$B_{\mu}$ following Ref.~\cite{Samoylov:2013xoa} in the form 
\begin{equation}
B_\mu(E_\nu) = \frac{B_\mu^{(0)}}{1+B_\mu^{(1)}/E_\nu}\, ,
\label{eq:bmu}
\end{equation}
which results in a rise of $B_\mu$ with $E_\nu$ at small $E_\nu$ and a 
saturation at large $E_\nu$. 

Since the fixed-target kinematics and the available beam energies 
do not allow for a coverage of the asymptotic region $\xi \gtrsim 50$ (Section~\ref{sec:nnlo}), 
we use only the NLO approximation in the QCD analysis of the charm di-muon data 
in (anti)neutrino CC DIS.   
In all our fits to NuTeV, CCFR and NOMAD data we 
use the nuclear corrections for the Fe target following 
Refs.~\cite{Kulagin:2004ie,Kulagin:2007ju,Kulagin:2010gd}.   
In order to reduce the computational time in our fits we do not apply electroweak 
corrections~\cite{Arbuzov:2004zr} to the NOMAD data, 
after verifying that they largely cancel out in the ${\cal R}_{\mu \mu}$ ratio.

\subsection{Inclusive charm production in $\nu$-emulsion interactions}
\label{sec:CHORUS} 

Experiments using nuclear emulsions can directly detect the individual charmed hadrons 
$D^0, D^+, D_s, \Lambda_c$ (and heavier charmed baryons) produced in (anti)neutrino interactions, 
through the location of the corresponding decay vertex. The information provided by emulsion experiments 
has the advantage that it does not rely upon semi-leptonic decays and it is therefore independent from $B_\mu$. 
The limitations of this type of measurement are mainly related to the low statistics available 
due to the small mass of nuclear emulsions usable in practice. 
The average nuclear target for (anti)neutrinos in nuclear emulsions is $A=81$ and $Z=36$.  
Only two data samples are currently available. 
The E531 experiment~\cite{Ushida:1988rt} collected 120 $\nu$-induced inclusive charm events 
and measured the complete decay and event kinematics, allowing for a determination of the charm 
production fractions, $f_h(E_\nu)$. 
The recent measurement by the CHORUS experiment~\cite{KayisTopaksu:2011mx} 
collected a total of 2013 inclusive charm events, although only the visible neutrino energy and 
the charm decay length were measured. The CHORUS experiment has also measured separately the yields of 
neutral ($D^0$) and charged charmed hadrons as a function of the neutrino energy.  

In this paper we focus on the ratio ${\cal R}_{c}\equiv\sigma_c/\sigma_{\rm CC}$ 
between the total charm cross-section and the inclusive CC 
cross-section as a function of the neutrino energy, 
which was published by the CHORUS experiment~\cite{KayisTopaksu:2011mx}.  
This ratio has the advantage of largely canceling out nuclear corrections related 
to the heavy nuclei in emulsions~\cite{Kulagin:2004ie,Kulagin:2007ju,Kulagin:2010gd}. 
It is worth noting that the direct charm detection in 
emulsions is potentially less sensitive to the details of the charm quark fragmentation than 
charm di-muon production. Since there is no exclusive selection of one particular decay mode and 
no fixed threshold on the momenta of the decay products, a larger phase space is detectable.  
The energy resolution achievable is, however, lower compared to electronic detectors like NOMAD. 
In order to be consistent with the measurement of ${\cal R}_{\mu \mu}$ from NOMAD and to 
have a reliable calculation of inclusive CC structure functions, 
we restrict our analysis to the kinematic region with $Q^2>1$ GeV$^2$.  
To this end, we directly evaluate the acceptance of this cut as a function of $E_\nu$ 
with the high resolution NOMAD data and apply the corresponding correction to the 
CHORUS measurement of ${\cal R}_c(E_\nu)$. We note that this acceptance 
correction is small (typically a few percent) and quickly vanishes with the increase of $E_\nu$.     

Since the typical momentum transfer is not too large compared to $m_c$, 
we use only the NLO approximation in the QCD analysis of CHORUS charm data.
In all our fits to CHORUS data we 
use the nuclear corrections for the average emulsion target following 
Refs.~\cite{Kulagin:2004ie,Kulagin:2007ju,Kulagin:2010gd}.   
Similarly to the case of NOMAD data discussed in Sec.~\ref{sec:NOMAD}, we do not apply electroweak 
corrections~\cite{Arbuzov:2004zr} to the CHORUS data, 
after verifying that they largely cancel out in the ${\cal R}_c$ ratio.  

Following the approach of Ref.~\cite{Alekhin:2008mb}, in all fits including charm di-muon data 
we constrain the inclusive semi-leptonic branching ratio $B_\mu$ with 
the measurement of charm production fractions $f_h(E_\mu)$ from the E531 experiment. These latter 
are convoluted with the recent values of exclusive branching ratios for charmed hadrons to 
determine $B_\mu$ at different neutrino energies, as shown in Figure~\ref{fig:bmu}.  
In addition, we use the recent direct 
measurement of $B_\mu$ in nuclear emulsions by the CHORUS experiment~\cite{KayisTopaksu:2005je}.  
As seen in Figure~\ref{fig:bmu}, the emulsion data from E531 and CHORUS are in agreement with a 
rising $B_\mu$ according to Eq.~(\ref{eq:bmu}).

\subsection{Associated W + charm production at the LHC}

The associated production of $W$-bosons and charm quarks in proton-(anti)proton collisions
at the LHC is directly sensitive to the strange parton distributions through the Born-level 
scattering off a gluon,
\begin{equation}
\label{eq:Wcborn}
g + s \to W + c\, ,
\end{equation}
and was proposed for a study of the strange distribution and asymmetry 
earlier~\cite{Berger:1988tu,Baur:1993zd,Lai:2007dq}. 
The LHC measurements of associated production of $W$-bosons and charm quarks probe the strange quark distribution 
in the kinematic region of $x \approx 0.012$ at the scale $Q^2=M_W^2$. The cross section of the 
associated $W$+charm production in proton-proton collisions at 
the LHC at a center-of mass of $\sqrt{s}$=7~TeV has been measured recently by the CMS~\cite{Chatrchyan:2013uja} 
and ATLAS~\cite{Aad:2014xca} collaborations. Both data sets are collected at $\sqrt{s}$=7~TeV and correspond to 
the integrated luminosity of 5 fb$^{-1}$. 
The $W$-boson candidates are identified by their decays into a charged lepton (muon or electron) 
and a neutrino, while the charm quark is tagged using hadronic and inclusive semi-leptonic decays of charm hadrons. 
The $W$-boson and the charm quark are required to have opposite charges. The same-charge combinations are 
subtracted to suppress potential contributions of the gluon splitting into a heavy-quark pair. The cross sections and 
cross-section ratios of $W$+charm production are measured differentially as a function of the pseudo-rapidity 
of the electron or muon originating from the $W$-boson decay and are provided together with
the correlations of the systematic uncertainties for both measurements. The results of the QCD analysis 
presented here use the absolute differential cross sections of $W$+charm production, measured in bins of 
the pseudo-rapidity of the lepton from the $W$-decay.

The CMS measurements~\cite{Chatrchyan:2013uja} of $W$+charm used in the present analysis are obtained for the 
transverse momenta of the lepton from $W$-decay larger than 35 GeV. The cross sections of $W$+charm 
production at CMS are determined at the parton level and the theory
predictions for the CMS measurements used in the present analysis
are calculated at NLO by using the MCFM program~\cite{Campbell:1999ah,Campbell:2010ff}
interfaced to APPLGRID~\cite{Carli:2010rw}.
The ATLAS measurement~\cite{Aad:2014xca} of associated $W$+charm production is
performed at the hadron level taking the transverse 
momentum of the $W$-decay lepton greater than 20 GeV. The 
theoretical predictions for the ATLAS data are obtained using the aMC@NLO
simulation, which combines an NLO QCD matrix-element calculation with a 
parton-shower framework~\cite{Frederix:2011zi}. At the parton-level
results of aMC@NLO were found to be in a good agreement with the MCFM predictions~\cite{Campbell:2005bb}.
In either case the theoretical calculations cannot be performed interactively in the PDF fit, 
since they are quite time consuming. 
Instead, for the ATLAS and CMS $W$+charm measurements, we employ the same approach 
implemented in the ABM12 fit~\cite{Alekhin:2013nda} to the LHC data
on the Drell-Yan process. The time-consuming theoretical predictions are 
computed only once 
for all members of a given PDF set, which encodes the PDF uncertainties. The 
resulting grid is later used in the fit so that  
the predictions corresponding to the values of the fitted PDF parameters are
estimated by an interpolation among the grid entries. Thus, lengthy
computations are only necessary during the fit preparation stage, while the fit itself
runs quite fast.  

\renewcommand{\theequation}{\thesection.\arabic{equation}}
\setcounter{equation}{0}
\renewcommand{\thefigure}{\thesection.\arabic{figure}}
\setcounter{figure}{0}
\renewcommand{\thetable}{\thesection.\arabic{table}}
\setcounter{table}{0}
\section{Determination of the strange sea quark distributions}
\label{sec:results}

In the earlier ABM12 fit~\cite{Alekhin:2013nda} the strange sea 
has basically been constrained by the NuTeV/CCFR data on charm di-muon production in 
(anti)neutrino CC DIS~\cite{Goncharov:2001qe}. 
Meanwhile, the recent NOMAD charm di-muon data~\cite{Samoylov:2013xoa} 
and the CHORUS inclusive charm data~\cite{KayisTopaksu:2011mx} in neutrino CC interactions 
allow an improved strange sea determination. 
Additional constraints on the strange sea can be obtained from the first CMS and ATLAS data on 
the associated $W$-boson and $c$-quark production~\cite{Chatrchyan:2013uja,Aad:2014xca}.
In the present paper we consider several variants of the ABM12 fit~\cite{Alekhin:2013nda} 
with different combinations of the new data sets together with  
the NuTeV/CCFR data~\cite{Goncharov:2001qe}:
\begin{itemize}
\item [ ]{\bf NuTeV/CCFR\ +\ NOMAD} aimed to check the impact of the NOMAD 
data~\cite{Samoylov:2013xoa}

\item [ ]{\bf NuTeV/CCFR\ +\ CHORUS} -- the same for the CHORUS 
data~\cite{KayisTopaksu:2011mx}

\item [ ]{\bf NuTeV/CCFR\ +\ CMS} -- the same for the CMS
data~\cite{Chatrchyan:2013uja}.

\end{itemize}
We also consider the following variants of the fit: 
\begin{itemize}
\item [ ]{\bf NuTeV/CCFR\ +\ CMS\ +\ ATLAS} -- to allow comparison of the 
 $W$+charm data obtained by 
ATLAS~\cite{Aad:2014xca} with the CMS measurements~\cite{Chatrchyan:2013uja}

\item [ ]{\bf NuTeV/CCFR\ +\ NOMAD\ +\ CHORUS} to check the cumulative impact of
the (anti)neutrino-induced charm production 
 data~\cite{Goncharov:2001qe, Samoylov:2013xoa,KayisTopaksu:2011mx}

\item[ ]{\bf CHORUS\ +\ CMS\ +\ ATLAS} to check the cumulative impact of the data 
sets~\cite{KayisTopaksu:2011mx,Chatrchyan:2013uja,Aad:2014xca} independent from  
the semi-leptonic branching ratio $B_\mu$. 
\end{itemize}
These fits have been upgraded as compared to the ABM12 one in the following respects:
\begin{itemize}
\item The NNLO corrections to the massive Wilson 
coefficients of Sec.~\ref{sec:nnlo} are taken into account 
when computing the $c$-quark contribution to the inclusive CC DIS 
structure function for the HERA data kinematics~\cite{Aaron:2009aa}. 
These data cover the range of $Q^2=300 - 15000~{\rm GeV}^2$ and  
therefore the asymptotic corrections can be safely applied. 
The numerical impact of the NNLO terms is about $5\%$ at the smallest values of $x\sim 0.01$ covered by the HERA CC
DIS data and it falls off to negligible values at $x=O(0.1)$, cf. Fig.~\ref{fig:heracc}. 
The new NNLO correction leads to an improved description of the data,  
with a value of $\chi^2$ reduced by 6 units for 114 data points in the HERA CC DIS
subset used in our analysis. 
In contrast, the charm di-muon data from (anti)neutrino CC DIS populate only the region 
of $Q^2\lesssim 100~{\rm GeV}^2$. 
Therefore, the asymptotic NNLO corrections cannot be applied to most of this kinematical range. 
Furthermore, due to the relatively small beam energy the highest  
values of $Q^2$ covered by (anti)neutrino data correspond to the values of $x=O(0.1)$, where the NNLO
corrections can be neglected. 

\item The inclusive semi-leptonic branching ratio $B_\mu$ for charmed hadrons is parameterized
according to Eq.~(\ref{eq:bmu}) to take into account the dependence on the neutrino  
energy $E_\nu$, rather than using a constant $B_\mu$ as in the earlier ABM fits. 
The parameters $B_\mu^{(0,1)}$ are fitted to the data simultaneously with 
the PDFs, high twist terms, strong coupling constant, mass of the charm quark etc. 
(cf. Ref.~\cite{Alekhin:2012ig} for the full list of fitted parameters). 
The large-$E_\nu$ asymptotic coefficient $B_\mu^{(0)}$ is partly constrained by the combination 
of neutrino- and antineutrino-induced charm di-muon data from NuTeV/CCFR  
like in the earlier ABM fits (cf. Ref.~\cite{Alekhin:2008mb} for details), 
while the coefficient $B_\mu^{(1)}$ is basically determined by the E-531
data~\cite{Ushida:1988rt} on $B_\mu$ and by the NOMAD charm di-muon data at small $E_\nu$ 
as in Ref.~\cite{Samoylov:2013xoa}.  
Our best estimate for $B_\mu(E_\nu)$ obtained in the variant of our analysis with the NuTeV/CCFR, NOMAD, and CHORUS
data included is displayed in Fig.~\ref{fig:bmu}. 
At large $E_\nu$ the shape of $B_\mu$ is comparable to the behavior taken for the ABM12 fit. 
The value of the coefficient $B_\mu^{(0)}=0.0933(25)$ obtained in the present
analysis is consistent with the $E_\nu$-independent determination $B_\mu=0.0904(33)$ in
ABM12~\cite{Alekhin:2012ig}. 
At small $E_\nu$ the value of $B_\mu$ falls off significantly 
and the coefficient controlling this behavior is determined in our analysis as $B_\mu^{(1)}=5.6\pm1.1~{\rm GeV}$.

\item As a minor improvement we also have corrected the absolute normalization
 of the ATLAS data on $W$- and $Z$-boson production~\cite{Aad:2011dm}, 
in line with the findings of Ref.~\cite{Aad:2013ucp}. 
However, this correction causes only little
improvement in the value of $\chi^2$ for the ATLAS data and practically does not
affect the PDFs extracted (cf. {\it Note added in proof} in Ref.~\cite{Alekhin:2013nda}). 

\end{itemize} 

The NOMAD data pull the strange distribution somewhat down as compared to the NuTeV/CCFR determination, 
as seen in Fig.~\ref{fig:exp}. 
The effect is particularly significant at large $x$ due to a better coverage of the low-$E_\nu$ region in
the NOMAD data sample. 
Correspondingly, the uncertainty in the large-$x$ strange sea is reduced. 
The quality of the overall description of the NOMAD data is rather good,  
with a value of $\chi^2/NDP=49/48$, where $NDP$ is the number of data points. 
However, the ${\cal R}_{\mu \mu}$ distributions as a function of $E_\nu$ and 
the partonic center-of-mass energy $\sqrt{\hat{s}}$ show a worse agreement with the fit, cf. Fig.~\ref{fig:nomad}. 
Furthermore, the variants of the fit based on the individual $E_\nu$- and $\sqrt{\hat{s}}$-distributions only 
exhibit some deviations from the fit in which all NOMAD data are included. 
In any case, the deviations observed are within the fit uncertainty and the PDFs obtained 
using the different NOMAD data subsets are consistent. 

The CHORUS data pull the strange distribution somewhat up in the entire range
of $x$, as visible in Fig.~\ref{fig:exp}. This is in contrast with the impact of the
NOMAD sample. However, both results are consistent within the uncertainties. 
In the variant of the fit including both NOMAD and CHORUS data these opposite trends
compensate each other so that the central value of the resulting strange sea 
distribution is close to the one preferred by the NuTeV and
CCFR data, cf. Fig.~\ref{fig:comb}. 
At the same time the error in the strange sea is improved, in particular at $x=O(0.1)$. 
The CHORUS data somewhat overshoot the fit, especially if the NOMAD data are included, 
cf. Fig.~\ref{fig:chorus}. 
In all variants of the fit the value of $\chi^2$ for the CHORUS data is within the 
range of $5 - 9$ for $NDP=6$, which is statistically acceptable. 

The CMS data on the associated $W$+charm production also prefer a somewhat
enhanced strange sea, cf. Fig.~\ref{fig:comb}. 
The absolute cross section measurements are much more sensitive to
the strange sea than the ratio of the individual $W^+ \bar{c}$ and $W^- {c}$ channels,
which is basically driven by the $s-\bar s$ PDF asymmetry. 
However, in both cases the experimental errors are much
bigger than the PDF uncertainties in the predictions based on the NuTeV/CCFR data. 
As a result, the variant of the fit with the NuTeV/CCFR and CMS data 
included does not deviate much from the ABM12 one, as shown in Fig.~\ref{fig:cms}.
Moreover, in this case the relative change in the strange sea due to the CMS data 
is only at the level of few percent, due to the constraint coming from the
NuTeV/CCFR sample, cf. Fig.~\ref{fig:charm}. If we release the constraint from the 
NuTeV/CCFR data, we can obtain a somewhat enhanced strange sea distribution.  
In particular, this trend is observed in the variant of fit based on the
combination of the CMS and CHORUS data only, cf. Fig.~\ref{fig:cms}.
In this case the low-$x$ asymptotic behavior of the strange sea is poorly 
determined. In order to improve the stability of the fit  
we impose an additional constraint on the low-$x$ strangeness exponent 
\begin{equation}
\label{eq:lowx}
a_{s}=-0.234\pm0.036\, ,
\end{equation}
resulting from the fit based on the combination of all (anti)neutrino data from 
NuTeV/CCFR, NOMAD, and CHORUS. 
The strange sea distribution obtained in this way is somewhat enhanced as compared to 
the ABM12 one, while the calculation goes essentially through the CMS data points.
The ATLAS data on $W$+charm production~\cite{Aad:2014xca} are also in good agreement
with this variant of the fit, cf. Figs.~\ref{fig:atlas1}-\ref{fig:atlas3}.
This fact demonstrates a good consistency between the CMS and ATLAS
measurements. A certain discrepancy is observed for
the ATLAS data points with the largest pseudo-rapidity of the $W$-decay leptons, although 
it is not statistically significant. Indeed, in the variant of fit including also the
CHORUS and CMS data a value of $\chi^2/NDP=33/38$ is obtained 
for the full ATLAS $W$+charm sample, taking into account both the experimental 
correlated uncertainties and the theoretical error related to the modeling of the initial- and
final-state radiation. For comparison, a value of $\chi^2/NDP=17/32$ is obtained
if the ATLAS data with the largest pseudo-rapidity of the $W$-decay are rejected. 
In the former variant the strange sea is enhanced within $1\sigma$ at 
$x\gtrsim 0.1$, in line with the tension observed, as shown in Fig.~\ref{fig:comb}. 
At the same time the strange sea distribution obtained in the variant with
the points at the largest pseudo-rapidity removed is in very good agreement with the
determination based on the CHORUS and CMS data only. 

A combination of the CMS and ATLAS $W$+charm data with the
CHORUS measurement defines the upper limit for the strange sea distribution which can be obtained in
our analysis, since these three samples prefer an enhanced strange sea compared 
to the one obtained in the ABM12 fit. 
We obtained this upper limit by including a combination of the CMS,
ATLAS, and CHORUS data into the fit, without the charm di-muon data from NuTeV/CCFR and
NOMAD, which are sensitive to the semileptonic branching ratio $B_\mu$. 
By imposing the low-x strange sea exponent constraint from Eq.~(\ref{eq:lowx}) in this fit, 
the small-$x$ strange sea distribution is determined as well as in the ABM12 fit, cf. Fig.~\ref{fig:comb}. 
In general, the resulting strange sea distribution is shifted upwards 
by some 20\% as compared to the fit based on 
the combination of the charm di-muon data from NuTeV/CCFR and NOMAD. 
At large $x$ this shift is statistically insignificant due to the large uncertainties, 
however at $x\sim 0.1$ it amounts to up to $2-3$ standard deviations. 
These numbers provide a bound on the outermost discrepancy 
in the strange sea determination preferred by different data sets considered 
since the NuTeV/CCFR and NOMAD pull the strange sea somewhat down as compared to ATLAS, CMS, and CHORUS. 
It is also worth noting that the impact of the combination of the
NOMAD and CHORUS data is much smaller and does not exceed the strange sea 
uncertainties, cf. Fig.~\ref{fig:comb}. 
We do not consider to add the ATLAS and CMS data to our final reference fit in view of the missing NNLO
QCD corrections to the hadro-production of $W$+charm. 
This choice does not lead to any essential change in the
strange sea distribution because of the rather big uncertainties in those data.

\section {Comparison with earlier determinations}
\label{sec:discussion}

The strange sea obtained in the variant of our analysis based on the (anti)neutrino induced 
charm production data from NuTeV/CCFR, NOMAD, and CHORUS is in agreement with the ABM12 one 
within the errors, cf. Fig.~\ref{fig:comb}. At the same time, the errors at
$x\gtrsim 0.01$ are largely improved, particularly at $x={\cal O}(0.1)$, where
the improvement in the error amounts to a factor of two. 
Conventionally, the magnitude of the strange sea is often presented in terms of an integral 
strangeness suppression factor 
\begin{equation}
\label{eq:kappas}
\kappa_s(\mu^2) \,=\, 
\frac{\int\limits_0^1x[s(x,\mu^2)+\bar{s}(x,\mu^2)]dx}{\int\limits_0^1x[\bar{u}(x,\mu^2)+\bar{d}(x,\mu^2)]dx}\, , 
\end{equation}
where $s$, $\bar s$, $\bar u$, and $\bar d$ are the strange, anti-strange, anti-up, and anti-down quark distributions, respectively. 
The value of $\kappa_s$ obtained in the variant of the present analysis including the NuTeV/CCFR, NOMAD, and CHORUS data
is comparable to the NOMAD~\cite{Samoylov:2013xoa} and CMS~\cite{Chatrchyan:2013mza} determinations, cf. Table~\ref{tab:kappas}. 
However, the error in $\kappa_s$ obtained by CMS is quite large due to the PDF parametrisation uncertainty.
At the same time the error in $\kappa_s$ obtained by NOMAD is smaller than ours. 
This fact can be explained by the constraints imposed in the NOMAD analysis 
on the low-$x$ strange sea behavior, which is poorly determined by the those data alone.  
It is also worth noting that the normalization of $\kappa_s$ in Eq.~(\ref{eq:kappas}), 
i.e. the second Mellin moment of $\bar{u}+\bar{d}$, is not fixed by any sum rule, 
and is therefore itself subject to variations in any given analysis. 
\begin{table}[ht!]                                                                               
\renewcommand{\arraystretch}{1.3}                                                                
\begin{center}                                                                                   
{\small                                                                                          
\begin{tabular}{|c|c|c|c|}                                                                     
\hline
 & present analysis& NOMAD~\cite{Samoylov:2013xoa} & CMS~\cite{Chatrchyan:2013mza} 
\\
 &(NuTeV/CCFR+NOMAD+CHORUS) & & 
\\
\hline
$\kappa_s(20~{\rm GeV}^2)$ &
$0.654\pm0.030$
 &
$0.591\pm0.019$
 &
$0.52\pm0.17$
 \\
\hline
\end{tabular}
}
\caption{\small
  \label{tab:kappas}
  The integral strangeness suppression factor Eq.~(\ref{eq:kappas}) obtained in 
  the present analysis in comparison with the earlier determinations. 
}
\end{center}
\end{table}

The $x$ dependence of the strange sea distribution is not much different from the non-strange ones. 
In particular, the shape of the $x$-dependent strange sea suppression factor  
\begin{equation}
r_s(x,\mu^2) \, = \, \frac{s(x,\mu^2)+\bar{s}(x,\mu^2)}{2 \bar{d}(x,\mu^2)}
\, ,
\end{equation}
preferred by the combination of the NuTeV/CCFR, CHORUS, and NOMAD data,
assumes roughly a constant value over the entire $x$-range, cf. Fig.~\ref{fig:ssup}. 
This is in line with the earlier analysis~\cite{Alekhin:2008mb} and other global PDF 
fits~\cite{Gao:2013xoa,Ball:2012cx,Martin:2009iq}. 
The value of $r_s$ as obtained from the combination of the CHORUS and CMS data is somewhat enhanced
at $x={\cal O}(0.01)$, although it suffers from large uncertainties. 
As discussed above, this combination of data gives an upper limit for the
size of the strange sea distribution determined in our analysis. This determination 
is consistent with
the results obtained by CMS~\cite{Chatrchyan:2013mza} from the analysis of their own
data on the $W$+charm production in combination with the HERA DIS data~\cite{Aaron:2009aa}.  
However, a much higher value of $r_s=1.00^{+0.25}_{-0.28}$ was obtained at $x=0.023$ and 
$\mu^2=1.9~{\rm GeV}^2$ in the ATLAS $epWZ$-fit~\cite{Aad:2012sb} to 
a combination of the ATLAS data on the $W$- and $Z$-production~\cite{Aad:2011dm}
together with the HERA DIS data.
The non-strange sea obtained in Ref.~\cite{Aad:2012sb} also differs from ours in several aspects. 
In particular, we obtain a positive iso-spin asymmetry of the sea $x(\bar d- \bar u)$, 
as preferred by the FNAL-E-866 Drell-Yan data~\cite{Towell:2001nh} included into our
analysis, cf. Fig.~\ref{fig:udm}. 
Instead, the value of $x(\bar d- \bar u)$ obtained in Ref.~\cite{Aad:2012sb} is negative, 
implying that the strange sea
enhancement is achieved at the expense of a suppression of the $d$-quark distribution. 
We note that the same picture is actually observed in the analysis by the NNPDF
collaboration (version 2.3) based on collider data only~\cite{Ball:2012cx}. 
Since the HERA inclusive DIS data do not allow to disentangle the flavor species of the PDFs, 
these peculiarities may be attributed to the impact of the ATLAS data.  

Our fit is in good agreement with the ATLAS data sample despite the fact that the
strange sea is suppressed by a factor of roughly two in the region of the ATLAS kinematics. 
Indeed, we obtain a value of $\chi^2/NDP=34.5/30$ for the ATLAS data 
in the variant including the NuTeV/CCFR, NOMAD, and CHORUS data. 
This is well comparable to the value of $\chi^2/NDP=33.9/30$ obtained in the analysis of Ref.~\cite{Aad:2012sb}. 
Furthermore, our value of $r_s=0.56\pm 0.04$ at $x=0.023$ and
$\mu^2=1.9~{\rm GeV}^2$ is, in fact, in agreement with the ones of Refs.~\cite{Aad:2012sb,Ball:2012cx}
considering the large uncertainties of the latter. 
Therefore, in principle the difference between the central values may be explained by a limited
potential of the existing collider data for the flavor separation of the quark PDFs. 
We also point out that additional discrepancies with respect to the analyses of Refs.~\cite{Aad:2012sb,Ball:2012cx} 
may appear due to the different factorization scheme employed to describe the DIS $c$-quark
production. However, this topic deserves a separate study.  

The \msbar value of the charm quark mass $m_c(m_c)$
obtained with the NuTeV/CCFR, NOMAD, and CHORUS data included into the fit, 
\begin{equation}
m_c(m_c)=1.222\pm0.024~({\rm exp.})~{\rm GeV}
\, ,
\label{eq:mc}
\end{equation} 
is consistent with the one of the ABM12 fit~\cite{Alekhin:2013nda}.
However, the experimental uncertainty is slightly improved due to the impact of the newly added 
NOMAD and CHORUS data. 
The value in Eq.~(\ref{eq:mc}) is also in agreement with the earlier determinations based on the DIS 
data~\cite{Samoylov:2013xoa,Gao:2013wwa,Alekhin:2012vu,Alekhin:2012un,Abramowicz:1900rp}
and the world average~\cite{Beringer:1900zz}, which has a comparable accuracy.

\section {Summary}
\label{sec:conclusion}

A detailed flavor separation of PDFs in the nucleon has become an important ingredient  
to achieve precise QCD predictions for current collider experiments, as well as for precision studies of 
electroweak physics in (anti)neutrino interactions. 
Of the light quark flavor PDFs the strange quark has been subject to the least
number of constraints by experimental data.
Using new data sets from the CHORUS and NOMAD experiments on charm quark production in 
neutrino DIS interactions, as well as LHC data on exclusive $W$+charm production, 
a significant reduction of the uncertainties in the determination of the strange quark PDF 
has been achieved with the present paper.

The ABM fit of PDFs and of the strong coupling constant $\alpha_s$ has 
so far used mainly NuTeV/CCFR data on charm di-muon production in neutrino-nucleus DIS
to constrain $s$ and $\bar{s}$ in the proton. The study described in the present paper 
is based upon the ABM framework and has considered the impact of new recent data sets  
relevant for the determination of the strange sea distribution. 
As a base line, the fit to the combined data of NuTeV, CCFR, NOMAD and CHORUS
has been shown to lead to small upwards shifts in the strange sea distributions ${\cal O}(5\%)$, 
while the extreme case using only a combination of CMS, ATLAS and CHORUS data 
leads to an upwards shift ${\cal O}(20\%)$. 
This latter result can be considered as an upper limit allowed by existing data.
As an additional benefit, the energy dependence of the semi-leptonic branching
ratio $B_\mu$ of the charmed hadrons, relevant for all (anti)neutrino induced charm di-muon data,  
has been determined with the help of the new NOMAD data.
The resulting strange quark PDF has been employed to obtain predictions for the  
exclusive $W$+charm production at the LHC. Comparisons with the available data from
CMS and ATLAS demonstrated a good consistency.

The results of the present analysis on the strange quark PDF do not support the ATLAS 
claim of an enhanced strange sea obtained in the $epWZ$-fit. Similar conclusions can be 
drawn with respect to the findings of the NNPDF (version 2.3) PDF fit including only 
collider data and disregarding any fixed-target data. 
In scrutinizing those analyses we have shown that, effectively, 
the strange sea enhancement observed by both the NNPDF (version 2.3) fit and the ATLAS 
$epWZ$-fit is the result of a suppression of the $d$-quark distribution.
Such a suppression leads to an additional discrepancy for the isospin asymmetry of the sea $\bar{d}-\bar{u}$  
with respect to the E-866 Drell-Yan data. Apparently, a separation of the individual 
quark flavor PDFs in the proton based entirely on collider data has strong 
limitations given the current experimental and theoretical uncertainties.

Future developments in both theory and experimental measurements are needed to improve the determination 
of the strangeness content of the proton. 
On the theory side, the complete NNLO QCD correction for heavy-quark CC DIS,
i.e. not just in the asymptotic regime of large $Q^2/m_h^2$, will 
minimize residual uncertainties in the analysis of the charm production data in (anti)neutrino DIS interactions.
Likewise, for the process $pp \to W+c$ at the LHC some gain in accuracy is to
be expected from a complete computation of the NNLO QCD corrections, e.g., for
the differential distribution $d\sigma(W^++c)/d\eta_l$.
On the experimental side, a measurement of $d\sigma(W^++c)/d\eta_l$ needs an ${\cal O}(3\%)$ accuracy in
order to improve upon the current status in the strangeness determination.
For the ratio $\sigma(W^++c)/\sigma(W^-+c)$ an ${\cal O}(1\%)$ measurement is needed.
If such an improvement in precision is feasible, a determination of the $s-\bar{s}$ asymmetry could be possible. 
The existing charm production data in (anti)neutrino-nucleus interactions are limited by the available 
statistics and by the knowledge of the semi-leptonic branching ratio $B_\mu$. The next generation neutrino 
scattering measurements~\cite{Mishra:2008nx,Adams:2013qkq} can address both issues, allowing for a substantial 
improvement in the precision of both $s$ and $\bar s$.

\subsection*{Acknowledgments}
We would like to thank Andrea Vargas Trevi$\tilde{\rm n}$o for providing ABM12
predictions for the CMS data and Alexander Hasselhuhn for discussions.  

S.A, J.B., and S.M. are grateful to the Mainz Institute for Theoretical Physics (MITP) 
for its hospitality and its partial support during the completion of this work, 
J.B. also acknowledges support from Technische Universit\"at Dortmund.

This work was realized within the scope of the PROSA collaboration. 

This work has been supported in part by Helmholtz Association under contracts VH-HA-101 ({\it Alliance Physics at the Terascale}) and SO-072, by Deutsche Forschungsgemeinschaft in Sonderforschungs\-be\-reich/Transregio~9, 
by Bundesministerium f\"ur Bildung und Forschung through contract (05H12GU8),
by the European Commission through contract PITN-GA-2010-264564 ({\it
  LHCPhenoNet}), and PITN-GA-2012-316704 ({\it Higgstools}).  

{\footnotesize                                                          

}

\newpage

\renewcommand{\thefigure}{\thesection.\arabic{figure}}
\setcounter{figure}{0}
\setcounter{section}{4}

\begin{figure}[tbh]
\centerline{
  \includegraphics[width=16.0cm]{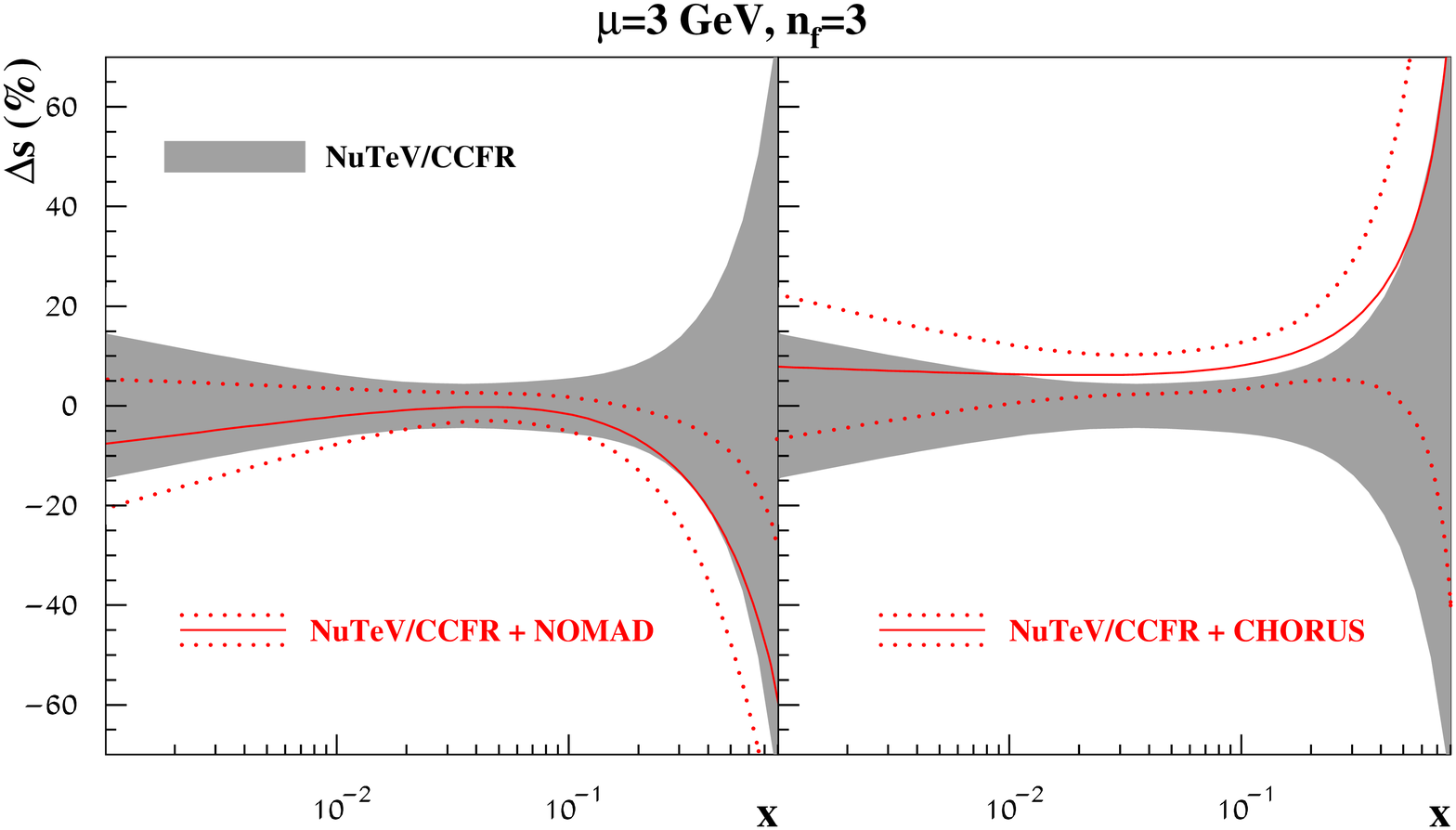}}
  \caption{\small
    \label{fig:exp}
      Relative strange sea uncertainty obtained from different variants of 
      the present analysis: the grey area represents the result with 
      only the NuTev and CCFR data~\cite{Goncharov:2001qe} 
      employed to constrain the strange sea, 
      the solid lines show 
      the relative change in the strange sea due to
      the NOMAD~\cite{Samoylov:2013xoa} (left panel) and 
      CHORUS~\cite{KayisTopaksu:2011mx} (right panel)
      data sets, respectively. The dotted lines correspond to
      the $1\sigma$ strange sea uncertainty after the inclusion of the new data sets. 
}
%
\vspace*{0.25cm}
%
\centerline{
  \includegraphics[width=15cm]{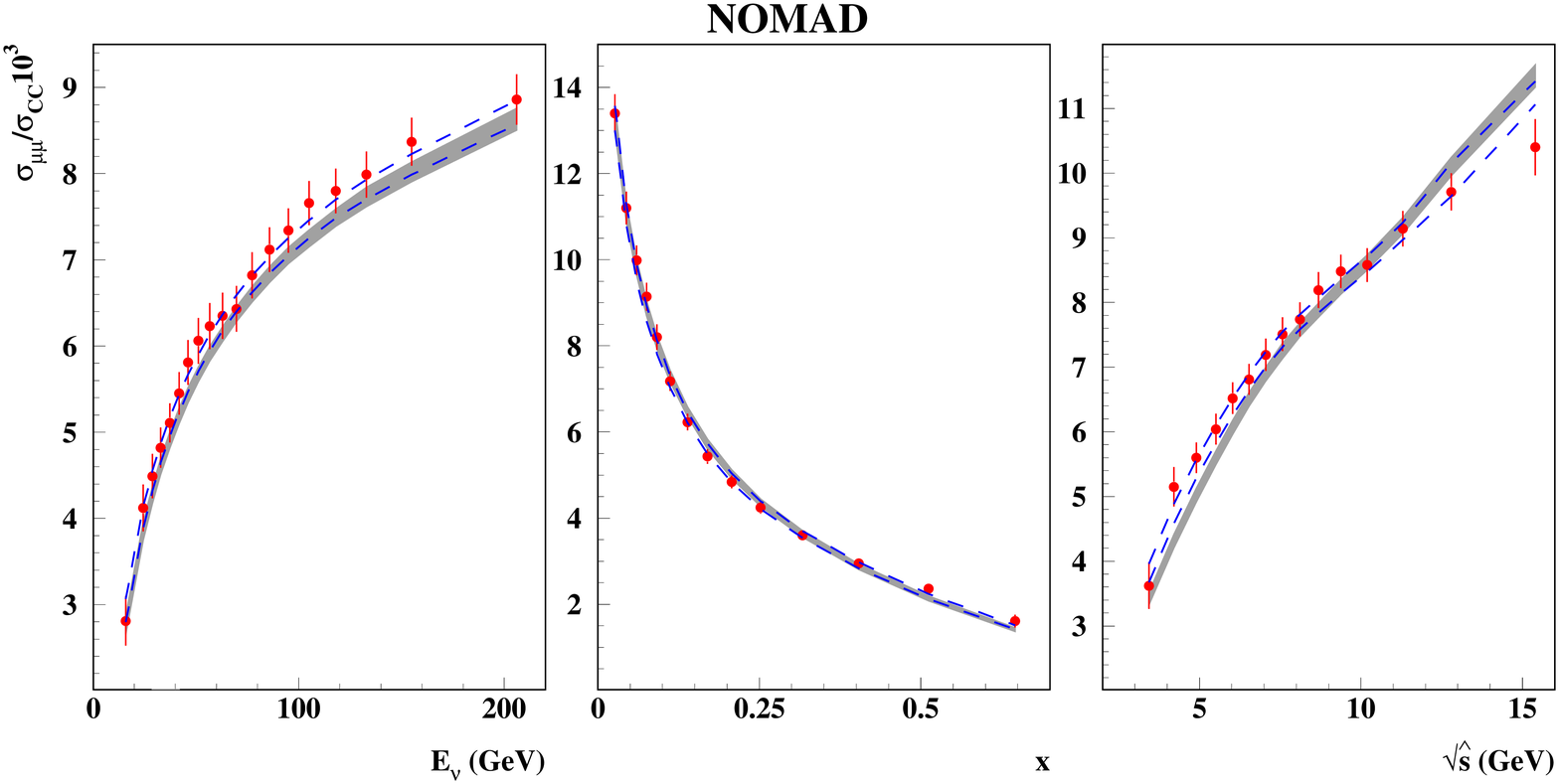}}
  \caption{\small
    \label{fig:nomad}
    Ratio ${\cal R}_{\mu \mu}$ between the cross-sections for charm di-muon production 
    and the inclusive CC neutrino-nucleon measured by the NOMAD
    collaboration~\cite{Samoylov:2013xoa} as a function of the beam 
    energy $E_\nu$ (left), the Bjorken $x$ (central), and the partonic 
    center-of-mass energy $\sqrt{\hat{s}}$ (right) 
    compared with the $1\sigma$ band obtained from the 
    variant of our fit with the NOMAD data included (shaded area). 
    The dashed lines display the $1\sigma$ band for the variants of 
    the fit based only on the respective NOMAD distributions.  
}
\end{figure}

\begin{figure}[tbh]
\centerline{
  \includegraphics[width=16.0cm]{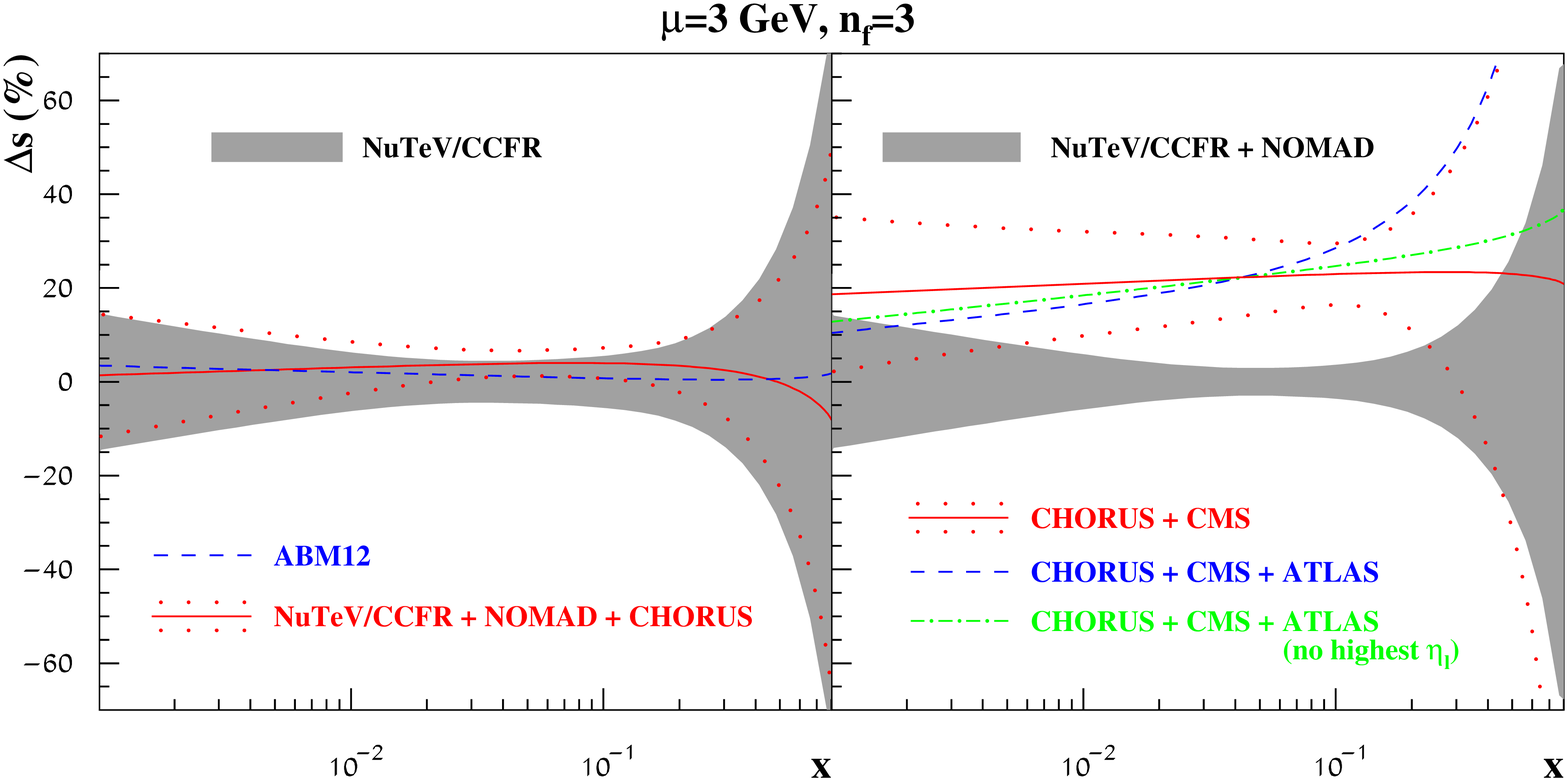}}
  \caption{\small
    \label{fig:comb}
Left panel: Same as Fig.~\ref{fig:exp} for the variant
of the present analysis with only the NuTeV/CCFR~\cite{Goncharov:2001qe}
data (grey area) in comparison with the one including also 
the NOMAD~\cite{Samoylov:2013xoa}
 and CHORUS~\cite{KayisTopaksu:2011mx} data; the dashed line displays the relative difference 
with respect to the ABM12 fit~\cite{Alekhin:2013nda}. Right panel: Same as the left panel for the variant of the
present analysis with the NuTeV/CCFR and NOMAD data in comparison
with the one including only the CHORUS and CMS~\cite{Chatrchyan:2013uja} data;
the relative changes in the
strange sea due to the addition of the complete set of the ATLAS $W$+charm data~\cite{Aad:2014xca}  
and the reduced set with the highest lepton pseudo-rapidity $\eta_l$ removed 
are also displayed as dashed and dotted-dashed lines, respectively.  
}
%
%
\centerline{
  \includegraphics[width=8.5cm]{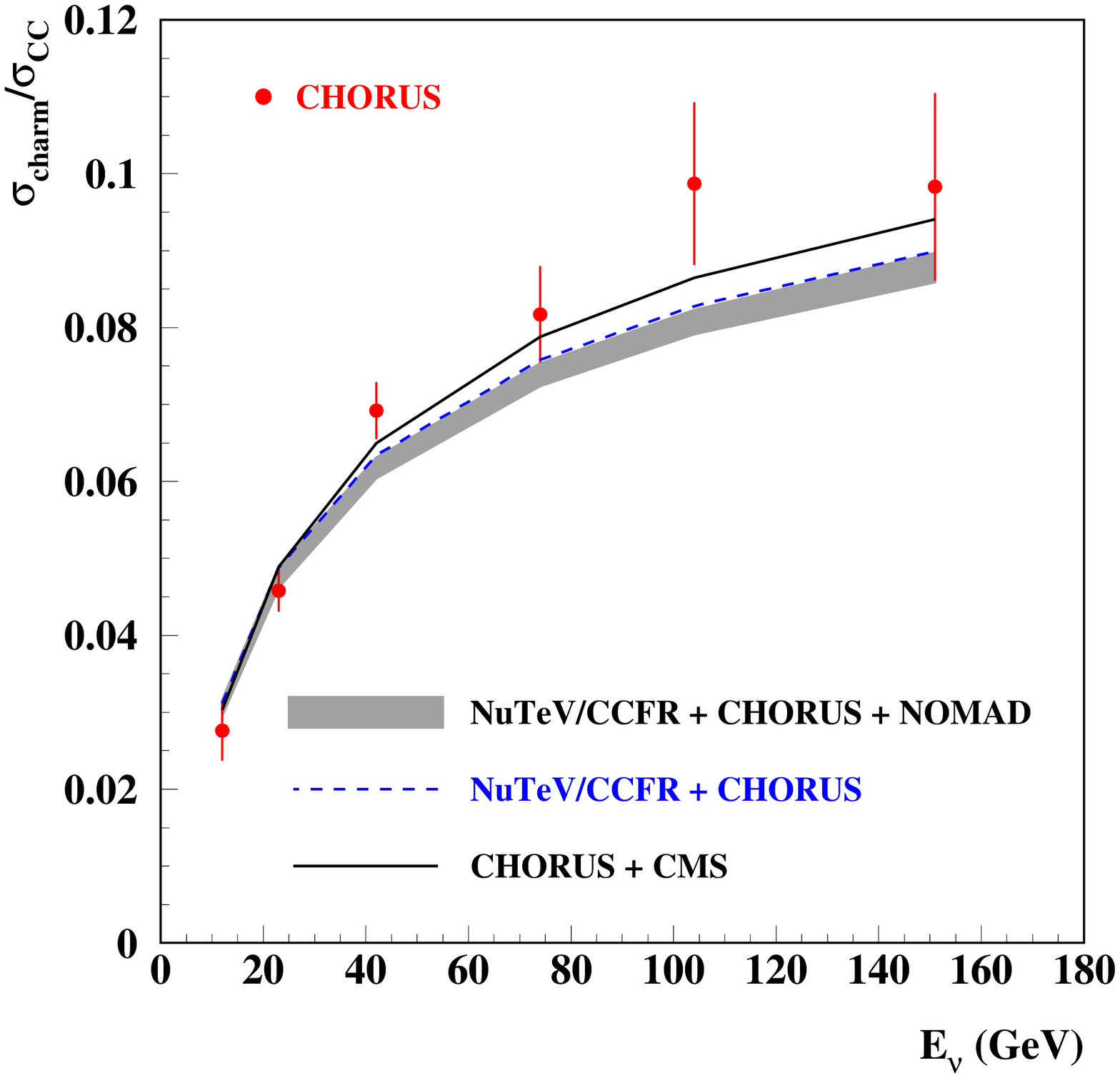}}
  \caption{\small
    \label{fig:chorus}
    Ratio ${\cal R}_c$ between the cross-sections for inclusive charmed hadron production 
    and the inclusive CC neutrino-nucleon measured by the CHORUS 
    collaboration~\cite{KayisTopaksu:2011mx} as a function of the beam energy $E_\nu$
    compared to the $1\sigma$ band obtained from the 
    variant of the present analysis with the
    NuTeV/CCFR~\cite{Goncharov:2001qe}, 
NOMAD~\cite{Samoylov:2013xoa}, and CHORUS
    data included (shaded area). The central values 
    corresponding to the variants with other combinations of the data sets used to 
    constrain the strange sea (solid line: CHORUS
   with CMS~\cite{Chatrchyan:2013uja} , 
    dashed line: NuTeV/CCFR with CHORUS) are also shown. 
}
\end{figure}

\begin{figure}[tbh]
\centerline{
  \includegraphics[width=16cm]{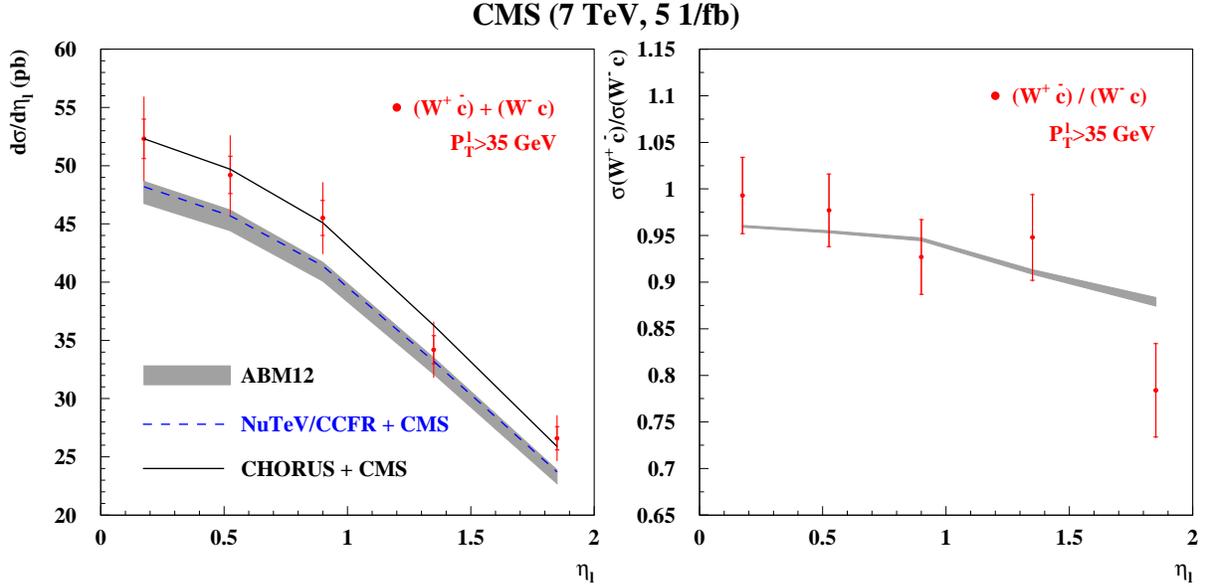}}
  \caption{\small
    \label{fig:cms}
    Same as Fig.~\ref{fig:chorus} for 
    the CMS cross-section data on the $W$+charm production~\cite{Chatrchyan:2013uja} 
    with a transverse momentum of the lepton from $W$-decay $P^l_T>35~{\rm GeV}$ 
    as a function of the lepton pseudo-rapidity $\eta_l$
    (left panel: sum of the absolute cross sections
    for the $W^+ c$ and $W^- \bar{c}$ channels, right panel: the ratio of 
    these two). The shaded area represents  
    the $1\sigma$ PDF uncertainty band from the ABM12 predictions.
}
\end{figure}

\begin{figure}[tbh]
\centerline{
  \includegraphics[width=16.0cm]{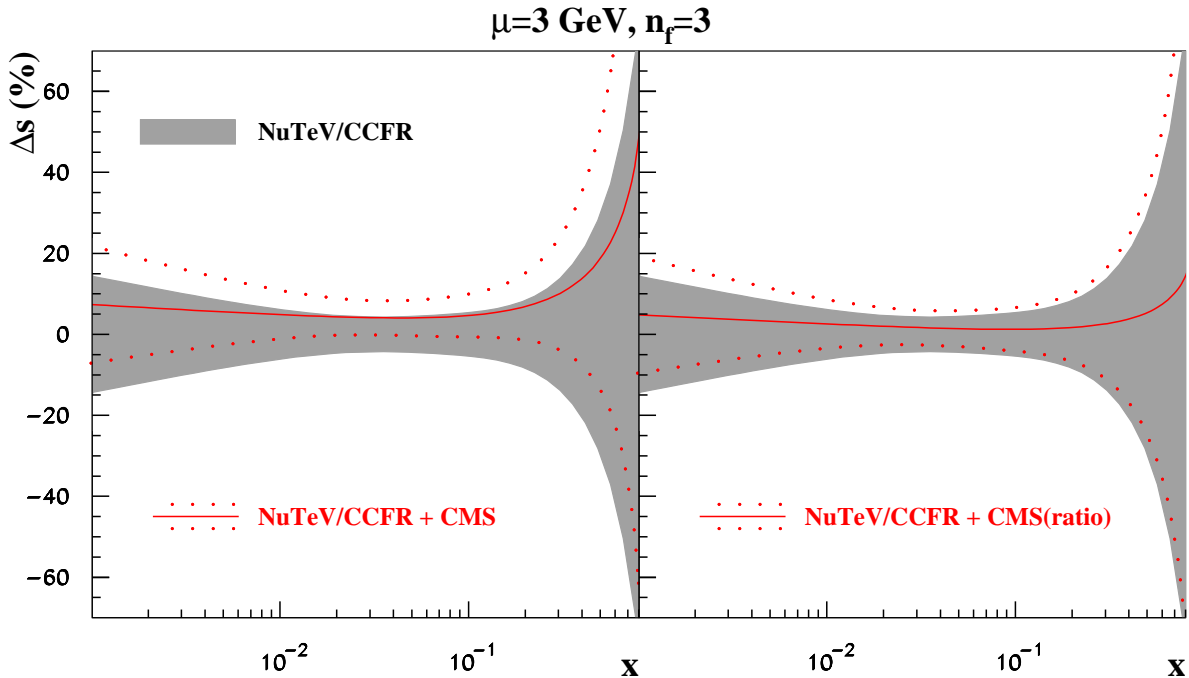}}
  \caption{\small
    \label{fig:charm}
      Same as Fig.~\ref{fig:exp} for the 
CMS data~\cite{Chatrchyan:2013uja} on the sum of 
the $W^-~c$ and $W^+~\bar{c}$ production cross sections
(left panel) and the ratio of these two (right panel). }
\end{figure}

\begin{figure}[tbh]
\centerline{
  \includegraphics[width=16.0cm]{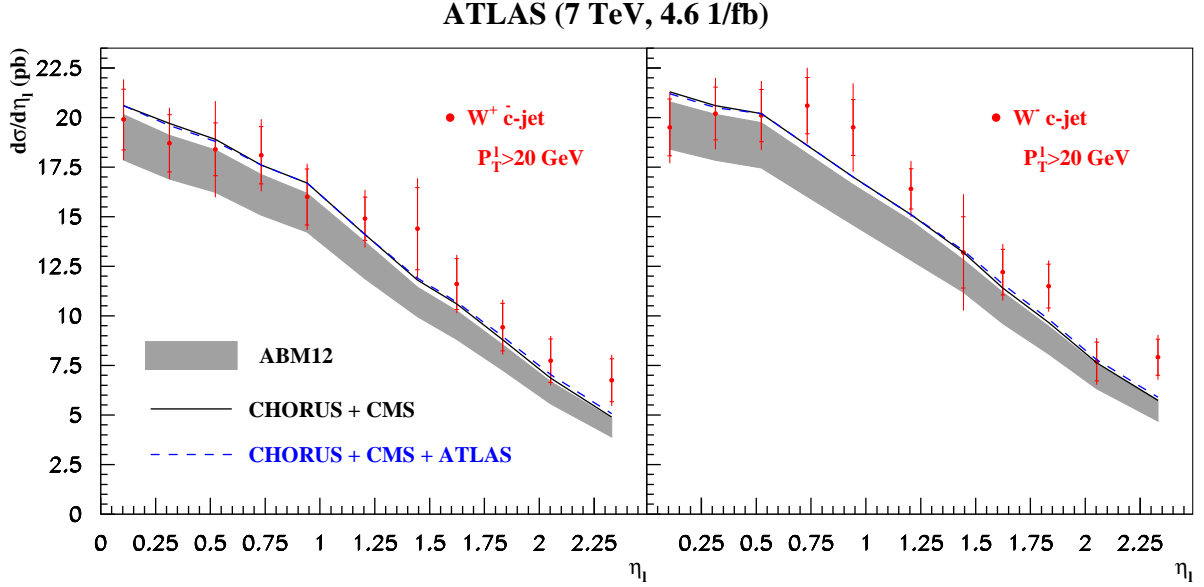}}
  \caption{\small
    \label{fig:atlas1}
      Same as Fig.~\ref{fig:cms} for the 
    ATLAS data on the cross section of the associated $W$-boson 
    and the charm jet production~\cite{Aad:2014xca}  
    with a transverse momentum of the lepton from $W$-decay $P^l_T>20~{\rm GeV}$ 
    as a function of the lepton pseudo-rapidity $\eta_l$
    (left panel: $W^+ c$-jet, right panel: $W^- \bar{c}$-jet). 
    The dashed line gives the central value of the present analysis with the 
    CHORUS~\cite{KayisTopaksu:2011mx}, 
CMS~\cite{Chatrchyan:2013uja}, and ATLAS~\cite{Aad:2014xca}
 data used to constrain the strange sea. The theoretical uncertainties related to
 the modeling of the initial- and final-state radiation are included into the
 ABM12 error band. 
}
\end{figure}

\begin{figure}[tbh]
\centerline{
  \includegraphics[width=16.0cm]{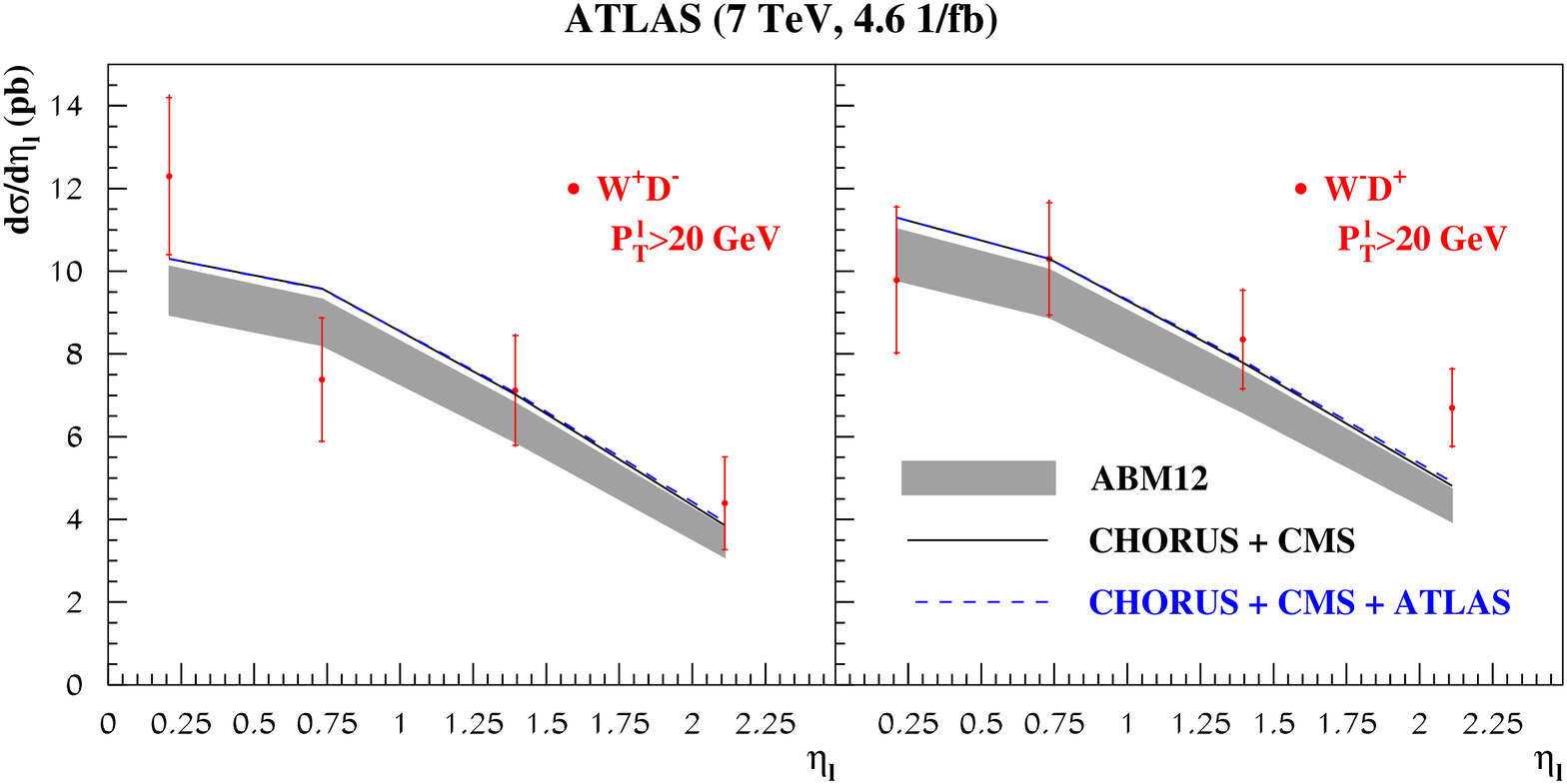}}
  \caption{\small
    \label{fig:atlas2}
      Same as Fig.~\ref{fig:atlas1} for the 
    ATLAS data on the cross section of the associated $W$-boson 
    and the $D$-meson production~\cite{Aad:2014xca}  
    (left panel: $W^+ D^-$, right panel: $W^- \bar{D}^+$).
}
\end{figure}

\begin{figure}[tbh]
\centerline{
  \includegraphics[width=16.0cm]{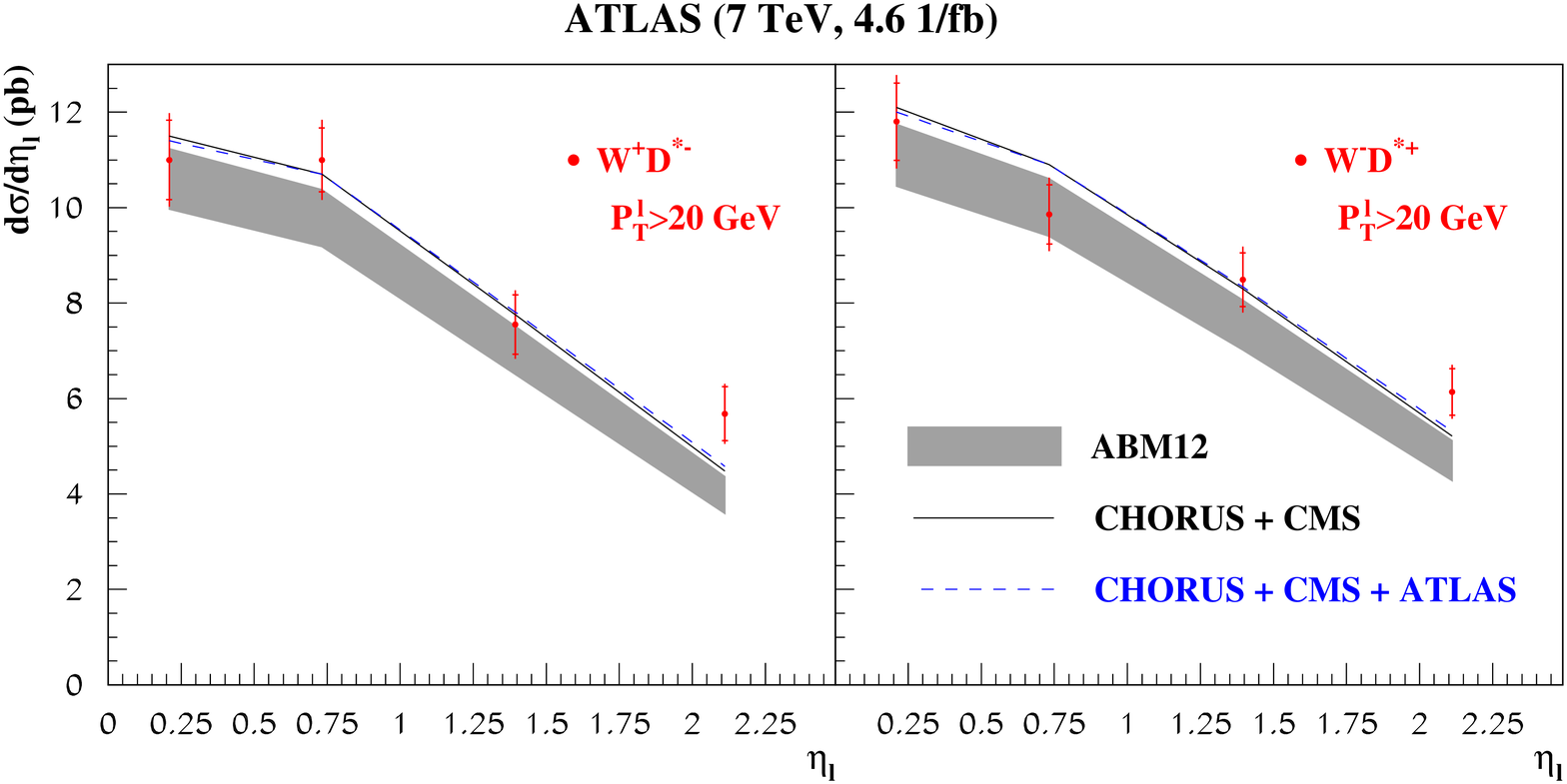}}
  \caption{\small
    \label{fig:atlas3}
      Same as Fig.~\ref{fig:atlas1} for the 
    ATLAS data on the cross section of the associated $W$-boson 
    and the $D^*$-meson production~\cite{Aad:2014xca}  
    (left panel: $W^+ D^{*-}$, right panel: $W^- \bar{D}^{*+}$).
}
\end{figure}

\renewcommand{\thefigure}{\thesection.\arabic{figure}}
\setcounter{figure}{0}
\setcounter{section}{5}

\begin{figure}[tbh]
\centerline{
  \includegraphics[width=12cm]{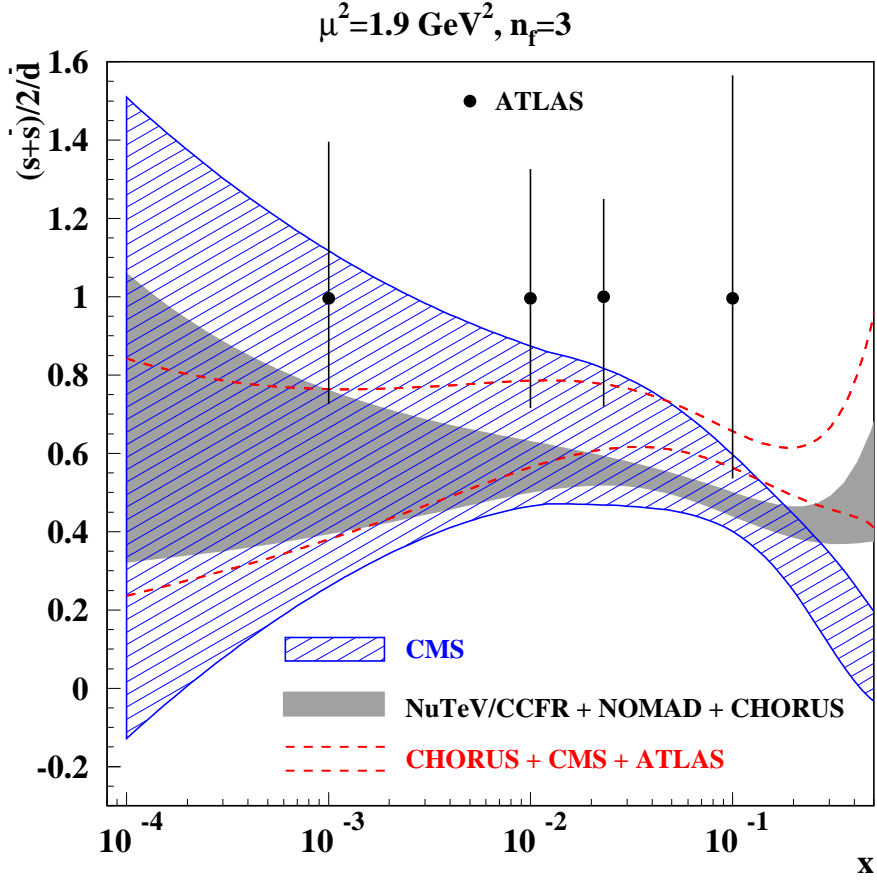}}
  \caption{\small
    \label{fig:ssup}
    The $1\sigma$ band for the 
    strange sea suppression factor $r_s=(s+\bar{s})/2/\bar{d}$ as a function of the 
    Bjorken $x$
      obtained in the variants of present analysis based on the 
    combination of the data by NuTeV/CCFR~\cite{Goncharov:2001qe},
CHORUS~\cite{KayisTopaksu:2011mx}, and NOMAD~\cite{Samoylov:2013xoa} 
    (shaded area) and CHORUS~\cite{KayisTopaksu:2011mx}, 
CMS~\cite{Chatrchyan:2013uja}, and ATLAS~\cite{Aad:2014xca} (dashed lines),  
in comparison with the results obtained by the CMS analysis~\cite{Chatrchyan:2013uja}
(hatched area) and by the ATLAS $epWZ$-fit~\cite{Aad:2012sb,Aad:2014xca} at 
different values of $x$ (full circles).
All quantities refer to the factorization scale $\mu^2=1.9~{\rm GeV}^2$. 
}
\end{figure}
\begin{figure}[tbh]
\centerline{
  \includegraphics[width=12cm]{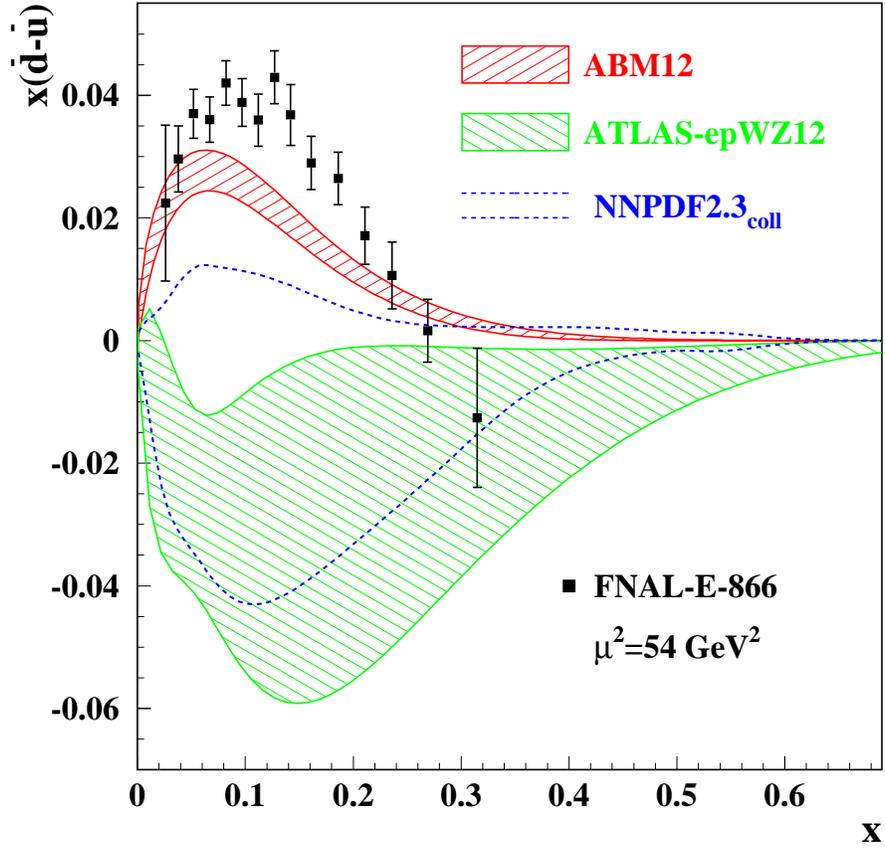}}
  \caption{\small
    \label{fig:udm}
    The $1\sigma$ band for 
    the iso-spin asymmetry of the sea $x(\bar{d}-\bar{u})$ at the scale of
    $\mu^2=54~{\rm GeV}^2$ as a function of the Bjorken $x$ obtained in the ABM12 fit 
    (right-tilted hatch), in comparison with the corresponding ones obtained by 
    the ATLAS~\cite{Aad:2012sb} (left-tilted hatch) and 
the NNPDF~\cite{Ball:2012cx} (dashed lines) analyses using only the LHC and HERA 
collider data. The values of $x(\bar{d}-\bar{u})$ extracted from the 
FNAL-E-866 data~\cite{Towell:2001nh} within the Born approximation are also shown as 
full circles with error bars.  
}
\end{figure}

\end{document}